\documentclass[10pt,twocolumn,final,3p]{elsarticle}

\usepackage[export]{adjustbox}
\usepackage{booktabs}
\usepackage{pdfpages}
\usepackage{hyperref}
\usepackage{blindtext}% for example text here only
\usepackage{color}
\usepackage{color,soul}
\usepackage{caption}
\usepackage{float}
\usepackage{xr-hyper} 
\usepackage{hyperref} 
\usepackage[utf8]{inputenc}
\usepackage{graphicx}
\usepackage{lipsum}
\usepackage[percent]{overpic}
\usepackage{stackengine}
\usepackage{subfig}
\usepackage{stfloats}
\usepackage{xspace} % for space after new commands
\usepackage{wrapfig}

\graphicspath{ {Images/} }

\journal{Scripta Materialia}

\begin{document}

\begin{frontmatter}
\title{Anisotropic diffusion of radiation-induced self-interstitial clusters in HCP zirconium: a molecular dynamics and rate-theory assessment} 

%%%%%%AUTHORS%%%%%%%%%%
\author[1]{Amir Ghorbani}
\author[1]{Yu Luo}
\author[1,2]{Peyman Saidi}
\author[1]{Laurent Karim Béland\corref{cor1}}

\ead{laurent.beland@queensu.ca}
\cortext[cor1]{Corresponding author}

\address[1]{Department of Mechanical and Materials Engineering, Queen's University, Kingston, ON K7L3N6, Canada}
\address[2]{Canadian Nuclear Laboratories, Chalk River, ON K0J 1J0, Canada}
\date{July 2023}

% \section{Abstract}
\begin{abstract}

Under irradiation, Zr and Zr alloys undergo growth in the absence of applied stress. This phenomenon is thought to be associated with the anisotropy of diffusion of either or both radiation-induced point defects and defect clusters. In this work, molecular dynamic simulations are used to study the anisotropy of diffusion of self-interstitial atom clusters. Both near-equilibrium clusters generated by aggregation of self-interstitial atoms and cascade-induced clusters were considered. The cascade-induced clusters display more anisotropy than their counterparts produced by aggregation. In addition to 1-dimensional diffusing clusters,  2-dimensional diffusing clusters were observed. Using our molecular dynamic simulations, the input parameters for the "self-interstitial atom cluster bias" rate-theory model were estimated. The radiation-induced growth strains predicted using this model are largely consistent with experiments, but are highly sensitive to the choice of interatomic interaction potential.

\end{abstract}

\end{frontmatter}

%%%%%%%%%%%%%%%%%%%%%MAIN TEXT%%%%%%%%%%%%%%%%%%%%%%%%%%
% \section{Main Body}

 HCP metals such as Mg, Ti and Zr -- and their alloys-- grow when exposed to high-energy particle irradiation.
 Radiation-induced growth (RIG) is volume conservative; it develops in the absence of applied stress and involves expansion in the prismatic (a) directions and contraction in the basal (c) axis. This transformation, unique to HCP materials, can be problematic in the context of nuclear power applications, since Zr alloys are widely used as fuel cladding materials in most commercial reactors and also as structural materials in CANDU (CANadian Deuterium Uranium) reactors due to their low neutron absorption cross-section \cite{choi2013radiation,holt2008reactor,woo1998modeling,christiaen2020influence,holt1988mechanisms,li2020cluster,christiaen2019new}. Despite decades of research, the physical nature of the micro-mechanisms explaining RIG remains unclear. 

Linking macroscopic RIG and microscopic defect evolution has been the topic of multiple theoretical efforts in the past 60 years \cite{pugh1963properties,woo1988theory,holt1993production,samolyuk2014analysis,barashev2015theoretical,golubov2014breakthrough}. Two leading theories, the Difference in Anisotropy of Diffusion (DAD) \cite{woo1988theory} model and the self-interstitial atom cluster bias (SIACB) \cite{golubov2014breakthrough} model are based on the assumption of anisotropic mass transport. Specifically, they both require that interstitial-type defects diffuse primarily along prismatic directions. This anisotropy in turn influences defect-defect and defect-sink interactions. Namely, it suppresses the growth of vacancy-type prismatic dislocation loops (a-loops) and interstitial-type c-loops and promotes the growth of interstitial-type a-loops and vacancy-type c-loops. The main difference between these models is that self-interstitial atoms (SIAs) are the agent of anisotropy in DAD while one-dimensional (1-D) gliding SIA clusters also play a major role in SIACB. Note that both models assume that c-loops appear once a certain radiation dose is reached, without providing micro-mechanistic explanations for this behaviour. The origin of vacancy c-loops formation has been the topic of numerous experimental and theoretical scientific studies \cite{christiaen2019new,liu2020two,dai2019mechanism,dai2021stability,harte2017effect}. Plausible explanations include the growth of small planar basal vacancy platelets into c-loops and cascade-induced collapse of multiple a-loops lying on the same basal plane into a c-loop \cite{dai2018atomistic}. 

Within the DAD framework, the SIA's diffusional anisotropy $D_a/D_c$ is assumed to be more than a hundred times larger than the vacancy's. However, \textit{ab initio} calculations by Samolyuk \textit{et al.} suggest that the difference in diffusional anisotropy
between SIAs and vacancies in Zr is not as large as that required by this model; in fact, the calculations suggest it may have the wrong sign \cite{samolyuk2014analysis}. In SIACB, large $D_a/D_c$ of self-interstitial defects is explained not by the anisotropy of diffusion of point defects, but rather by the anisotropy of diffusion of 1D gliding clusters. In this model, the growth rate is controlled by $\varepsilon_{i}^{g}$, the ratio of 1-D gliding SIA clusters to total SIAs, and $n_{i}^{g}$, the mean number of SIAs in these clusters.
Previous molecular modelling--using older force-fields--indicates that glissile SIA clusters can be produced by collision cascades in Zr \cite{voskoboinikov2006atomic,gao2001temperature,voskoboinikov2006identification}. Other modelling studies focused on the diffusion behaviour of individual sessile and glissile SIA clusters \cite{whiting1996computer,de2002mobility,de2011structure,march2022defect} and of clusters emerging from the aggregation of a large number of point defects \cite{maxwell2020atomistic}. Questions remain as to whether SIA clusters directly generated by collision cascades and aggregation of point defects present the same degree of diffusional anisotropy. Furthermore, the propensity for 1D gliding clusters generation by cascades has not been revisited using more recent force-fields.

In this article, we model and compare the diffusional anisotropy of SIA clusters produced by aggregation with those produced directly by collision cascades (see supplementary material (\href{run:./SM/SM.tex}{SM}) - sections 3 \& 4). First, we re-parametrized two widely used zirconium force-fields at short interatomic distances based on density functional theory (DFT) calculations. Second, the diffusion behaviour of SIA clusters produced by the aggregation of SIAs is presented. Third, the diffusion behaviour of SIA clusters produced in cascades is described.
Finally, estimates of $\varepsilon^{g}$ and $n^{g}$ are reported, and predictions pertaining to growth behaviour using a rate-theory model (i.e. a sensitivity analysis) are provided. 

The choice of force-field plays a crucial role in determining the mobility of SIA clusters \cite{march2022defect,de2011structure}. Furthermore, primary damage production is highly sensitive to the method chosen to model short-to-mid-distance (1 to 2 \AA) interatomic interactions \cite{stoller2016impact,beland2016atomistic,beland2017accurate,byggmastar2018effects,becquart2021modelling}.  Here, Mendelev and Ackland’s \cite{mendelev2007development} Zr potentials, referred to here as M2 and M3, were chosen and reparameterized at short interatomic distances following the procedure explained in Ref. \cite{stoller2016impact}. We refer to the reparametrized potentials as \href{run:/SM/M2R.eampot}{M2R} and \href{run:/SM/M3R.eampot}{M3R} (see \href{run:/SM/Models/20230414-SM.pdf}{SM - section 2}). Elastic constants and SIA formation energies were not materially affected by this reparametrization.

\begin{figure}[!t]
\centering
\includegraphics[width=1.2\linewidth]{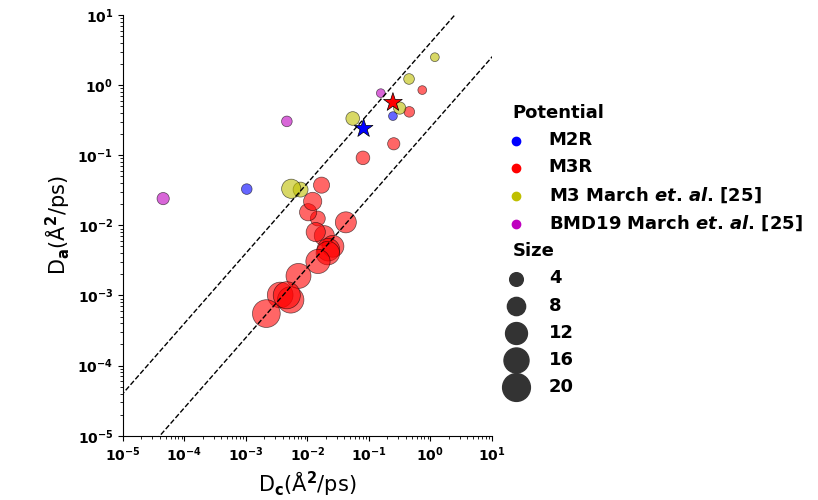}
\caption{Diffusion coefficients along the basal and prismatic planes of SIA clusters generated by aggregation of SIAs. Here, the circular marker radius is associated with the size of the corresponding SIA cluster, while the stars are single SIAs. Values within the dashed lines exhibit diffusional anisotropy corresponding to a bias factor of less than 25\%. We also display simulation results from March \textit{et. al.} \cite{march2022defect}, who employed the M3 and BMD19 potentials.}
\label{FIG:hist}
\end{figure}

Using the reparameterized potentials, we performed molecular dynamic (MD) simulations to study SIA clusters produced by aggregation of individual SIAs, comprised of up to 20 SIAs, to track their mobility. Each cluster was simulated for 100 ns at 573K in a box containing more than 15000 atoms using the Large-scale Atomic/Molecular Massively Parallel Simulator (LAMMPS)\cite{LAMMPS} with a timestep of 1 fs in the NVT ensemble employing a Nos\'{e}-Hoover thermostat. SIAs were added to the simulation cell one at a time, allowing them to form a single cluster before adding another (see \href{run:./SM/SM.tex}{SM - section 3} and \href{run:/SM/Videos/Video1.mp4}{video 1}). While M2R SIA clusters tend to form sessile platelets, M3R clusters tend to form glissile configurations.

Diffusion coefficients of these clusters along basal (D$_c$) and prismatic (D$_a$) planes were calculated and reported in Fig. \ref{FIG:hist} (see \href{run:./SM/SM.tex}{SM - section 3.2}). To distinguish between isotropic and anisotropic behaviour, we chose a D$_a$ to D$_c$ ratio threshold of four, which corresponds to a bias factor of 25\%, following the model proposed in Ref. \cite{saidi2020method} (see \href{run:./SM/SM.tex}{SM - section 5}). The dashed lines in Fig. \ref{FIG:hist} bound the diffusion coefficient in the isotropic regime. Most of the clusters diffuse roughly isotropically. Furthermore, diffusivity diminishes as clusters become larger \cite{maxwell2020atomistic}.
Instead of forming clusters by aggregation of SIAs, March \textit{et. al.} \cite{march2022defect} directly introduced crowdion SIA clusters in their simulation box, using the M3 and BMD19 potentials. Their results are included in the data presented in Fig. \ref{FIG:hist}. 

Irradiation of pure Zr was simulated in LAMMPS assuming primary knock-on atom (PKA) energies of 10, 20, 30 and 40 keV at 573K (see \href{run:./SM/SM.tex}{SM - section 4} and \href{run:/SM/Videos/Video2.mp4}{video 2}). 64 equidistributed initial PKA directions were chosen in one-sixth of a sphere for each potential and PKA energy. The simulation boxes contained more than 2.5 million atoms ($140\times80\times60$ unit cells) and time integration using a variable step size was performed for a duration of 500 ps. No thermostat was applied except for an exterior layer of 2 unit cells where we applied a Langevin thermostat at 573K.

\begin{figure}[!t]
\centering
\includegraphics[width=0.99\linewidth]{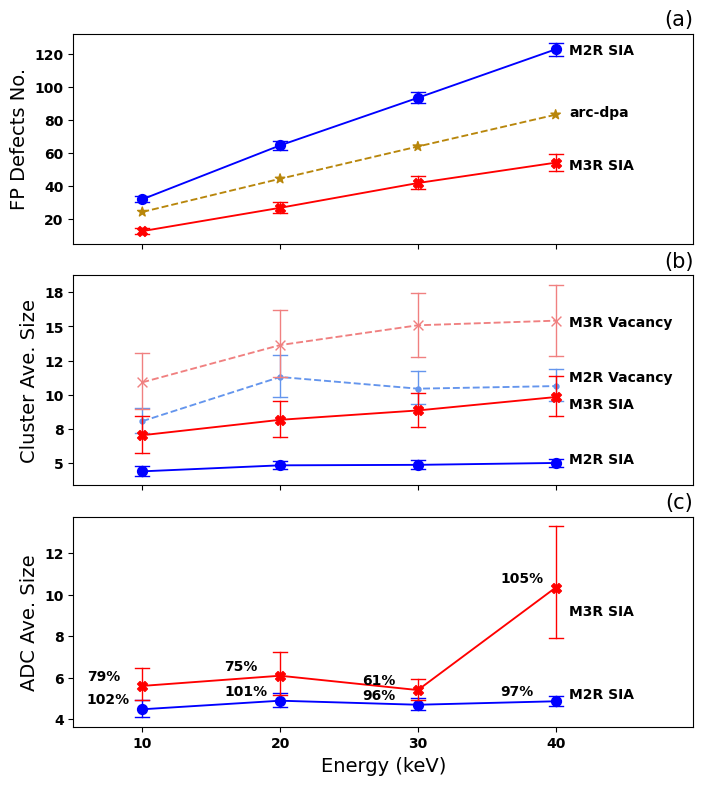}
\caption{Primary damage production in HCP Zr. In panel (a), we report the total number of displaced atoms as a function of the PKA kinetic energy. 64 MD runs using each force-field were averaged and compared to the predictions of the arc-dpa model. In panel (b), we report the average SIA and vacancy cluster size as a function of the PKA kinetic energy. In panel (c), we report the average size of ADCs (a-biased diffusing clusters) as a function of the PKA kinetic energy in comparison with the average SIA cluster size. Error bars correspond to confidence intervals of 95\%.}
\label{fig:cascades}
\end{figure}

Wigner-Seitz cell analysis (WS) was performed using OVITO \cite{ovito} to determine the number of Frenkel pairs (FP) present in the simulation cells (see \href{run:./SM/SM.tex}{SM - section 4.2}).
Fig. \ref{fig:cascades} (a) shows the averaged remaining FPs at the end of the cascade. For comparison, an estimate of FP production based on the  athermal recombination corrected displacement per atom (arc-dpa) model is also provided \cite{nordlund2018improving,yang2021full}. Note that this arc-dpa estimate is itself fitted on cascades simulated using the non-reparametrized M2 potential \cite{mendelev2012development,wilson2015anisotropy,yang2018molecular}. FP production increases in tandem with energy. Cascades simulated using the M2R potential led to a greater amount of FP produced than those performed using M3R. The arc-dpa estimates fall between these values.

Next, cluster analysis was performed using a cutoff distance of 5.8 \AA. The size of the SIA clusters remaining at the end of the cascade was averaged and illustrated in Fig. \ref{fig:cascades} (b). Greater PKA energy results in a slight increase in cluster size. The average SIA cluster size associated with M3R is double that associated with M2R.

To calculate $n_{i}^{g}$ and $\varepsilon_{i}^{g}$, we analyzed the behaviour of SIA clusters both visually and algorithmically to distinguish different cluster types. Less than 200 1-D gliders were observed in the 256 M2R cascade simulations; when considering M3R, this number is limited to 70. These 1-D clusters were mostly comprised of 6 to 11 SIAs (see \href{run:./SM/SM.tex}{SM - section 4.4.1}). 

\begin{figure}[!t]
\centering
\includegraphics[width=0.83\linewidth]{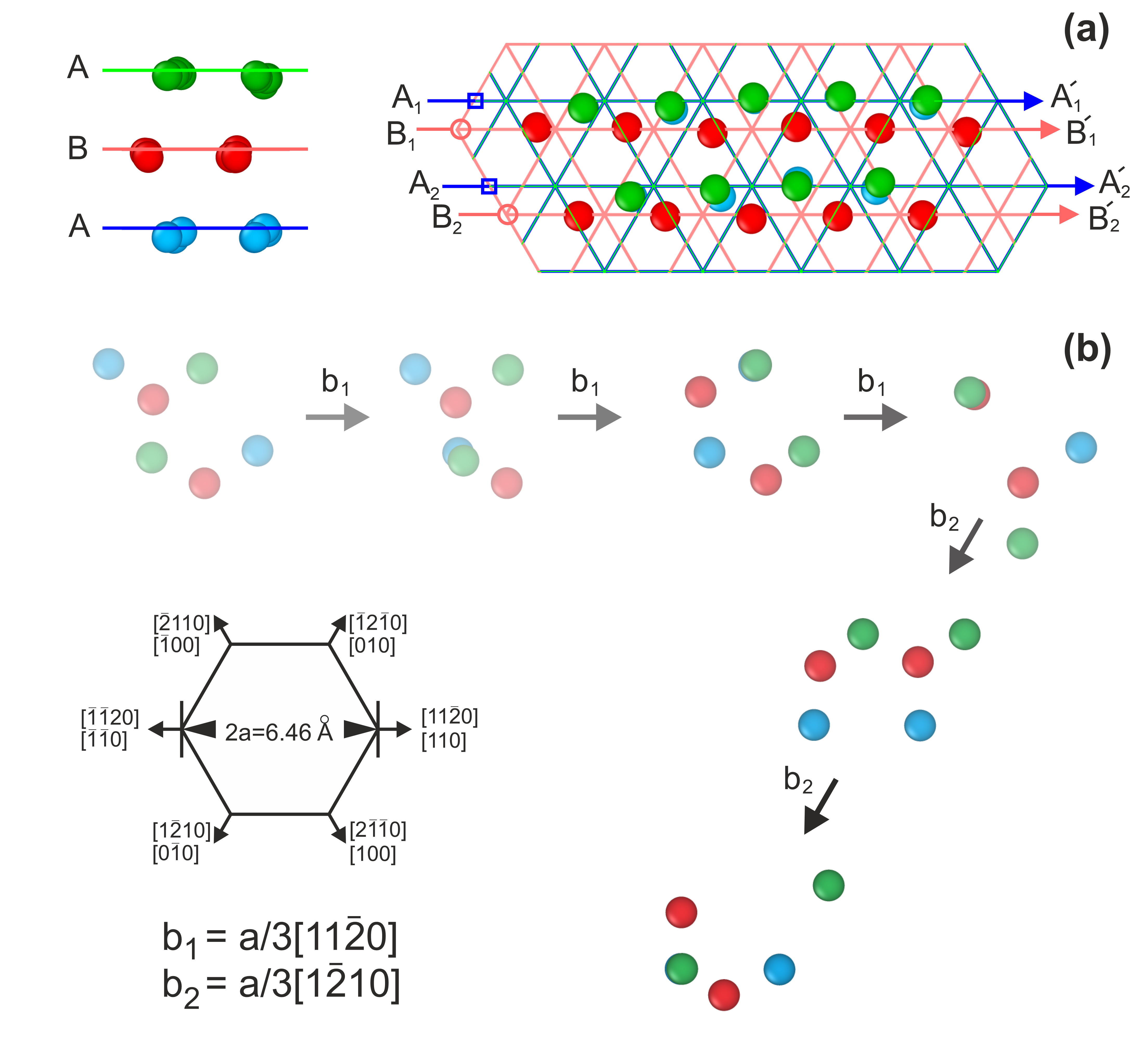}
\caption{Graphical representation of an ADC. (a) Equivalent sphere analysis is used to demonstrate the configuration of a perfect 1-D glider from prismatic (left) and basal (right) view. As a result, off-site atoms are illustrated and the Burgers vector is drawn using the structural mesh. Here, three consecutive basal planes of an ADC with 6 SIAs are identified by letters A (blue), B (red) and A (green) based on their position along the c axis. On each plane, 2 crowdions are identified along parallel directions, e.g. $\overrightarrow{A_1A_1^{\prime}\hspace{0.1cm}}$ and $\overrightarrow{A_2A_2^{\prime}\hspace{0.1cm}}$ on plane A. (b) Wigner-Seitz cell analysis is used to trace the gliding behaviour of a 2-D crowdion cluster containing 6 SIAs. Such SIA clusters can reconfigure themselves and change their direction of pure 1-D diffusion from one compact direction to another on the same plane.}
\label{fig:SphereAnalysis}
\end{figure}

Next, the 1-D clusters were visualized using equivalent sphere analysis (see \href{run:SM/20230414-SM.pdf}{SM - section 8} and \href{run:SM/Videos/Video3.mp4}{video 3}) \cite{terentyev2006effect}. The common feature of 1-D gliders is consecutive crowdions stacking along the c-axis and seamless arrangement along the a-axis. As shown in Fig. \ref{fig:SphereAnalysis} (a) left, six crowdion SIAs form three pairs in basal planes. These crowdions are shown along the a-axis in Fig. \ref{fig:SphereAnalysis} (a) right. The majority of the clusters show the same symmetric pattern with a different number of crowdions. These clusters move back on forth along [11$\bar{2}$0] direction, without a change in their nature. Another notable characteristic of the clusters, mainly observed in M3R interatomic potential, was the reconfiguration of crowdions to change the mobility direction from one a-axis to another. This behaviour is illustrated in Fig. \ref{fig:SphereAnalysis} (b). These are 2-dimensional (2-D) gliding clusters. Interestingly, such 2D motion of SIA clusters was reported in the past \cite{maxwell2020atomistic,de2002mobility}, but, to the best of our knowledge, their quantitative impact on RIG has not been assessed.  At the moment of transition, crowdions move out of the plane and rearrange to form crowdions in an alternative $<11\bar{2}0>$ direction. These are a-biased diffusing clusters, which should contribute to planar growth and shrinkage alongside 1-D gliding clusters.

\begin{figure}[!t]
\centering
\includegraphics[width=1.0\linewidth]{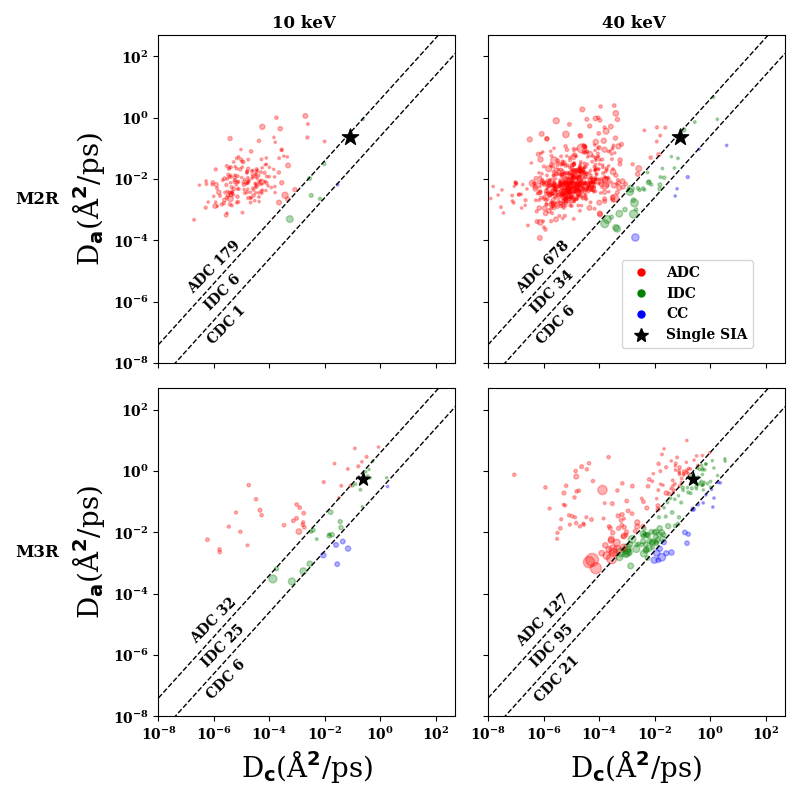}
\caption{Diffusion coefficients of cascade-induced SIA clusters. Calculations were performed using the M2R and M3R potentials. Two PKA energy of 10 and 40 keV are illustrated (20 and 30 keV results available in the S.I.). Each dot represents one SIA cluster and the star markers show the single SIAs. Data are color-coded based on the ratio of D$_a/$D$_c$.
The red colored points can be interpreted as ADCs, the green colored dots can be interpreted as roughly isotropically diffusing SIA clusters, and the blue colored dots can be interpreted as climbing clusters. The dashed lines bound the diffusion coefficients along the basal and prismatic planes where diffusion is considered to be isotropic (i.e. bias factor less than 25\%).}
\label{fig:DaDc}
\end{figure}

The diffusion coefficients (both along basal and prismatic planes) of each cascade-generated SIA cluster were estimated by tracking their size and position of their center of mass (see \href{run:./SM/SM.tex}{SM - section 4.4.2}). D$_c$ vs. D$_a$ is illustrated in Fig. \ref{fig:DaDc}. Clusters with a diffusion coefficient lesser than $D_{mob}=10^{-8}$ {\r{A}/ps} were considered immobile. Single SIAs are shown by stars and characterized by D$_a$ to D$_c$ ratio of 1.13 and 0.89 for M2R and M3R, respectively.
The diffusion coefficients of cascade-induced SIA clusters are higher than those of SIA clusters created by agglomeration, more remarkably for M3R (see \href{run:./SM/SM.tex}{SM - section 4.4.2}). This indicates defects created during extreme out-of-equilibrium conditions can differ substantially from those produced in near-equilibrium conditions \cite{calder2010origin}. Clusters created by the interaction of subcascade shockwaves have a higher propensity to lead to crowdion formation than the aggregation of point defects. Furthermore, the stress fields induced by the cascade debris may also play a role; such interactions are known to have an important effect on the kinetics of defects \cite{beland2013replenish,beland2013long,beland2015kinetic,brommer2014understanding,beland2015slow,beland2016differences,barashev2001reaction}.

\begin{figure}
\centering
\includegraphics[width=1.0\linewidth]{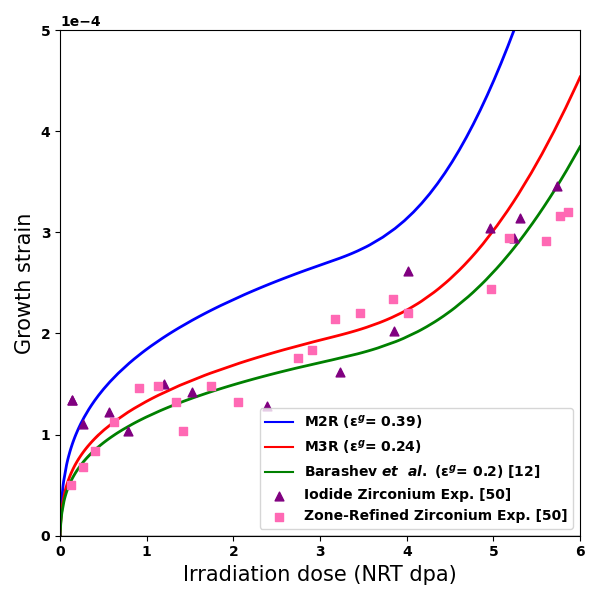}
\caption{Rate theory prediction of prismatic RIG in Zr using $\varepsilon^{g}$ calculated for the M2R- and M3R-based collision cascade simulations as compared to Barashev \textit{et al.}'s SIACB model and experimental results \cite{barashev2015theoretical,carpenter1988irradiation}. In this model, growth strain is measured against displacement per atom (DPA) in the procedure suggested by Norgett-Robinson-Torrens
model (NRT) \cite{NORGETT197550}.}
\label{fig:SIACB}
\end{figure}

Using the same bias factor threshold as previously, we classified SIA clusters into three groups: a-biased diffusing clusters (ADC), isotropically diffusing clusters (IDC), and c-biased diffusing clusters (CDC). SIACB model is based on the diffusion anisotropy of clusters, and their preferential interaction with different sinks. Therefore, in addition to 1-D gliders that were initially considered by the SIACB model, we include all ADCs to calculate $n^{g}$ and $\varepsilon^{g}$. As can be seen in Fig. \ref{fig:cascades} (c), $n^{g}$ is highly sensitive on the choice of force-field, but roughly independent of the PKA energy. The average size of ADCs is found to be in the same range of the cluster average size. The migration behavior of small clusters is known to be highly sensitive to the choice of force-field (see, e.g. Ref. \cite{march2022defect}). In particular, the anisotropy of diffusion of M2R di-SIAs (SIA clusters of size 2) is very large, but M3R di-SIAs are nearly isotropic. Furthermore, M2R cascades tend to generate many more di-SIAs than M3R cascades.
Hence, using M2R and accounting for di-SIAs would lead to a very large estimate of $\varepsilon^{g}$ ($\approx$0.62). However, both the large production rate of di-SIAs and their large anisotropy of diffusion may be spurious. For this reason, we opted to exclude di-SIAs clusters from the analysis presented below, which can be viewed as a conservative lower bound estimate of $\varepsilon^{g}$. Estimates including di-SIAs are available in the \href{run:./SM/SM.tex}{SM - section 6}.
Our M2R simulations are associated with $\varepsilon^{g}\approx$ 0.39, while M3R simulations are associated to 0.24. These values are based on an automated analysis of the SIA clusters; visual analysis leads to a value of 0.2. This compares to about 0.25 as estimated in Ref. \cite{voskoboinikov2006atomic} using an older force-field \cite{ackland1995defect}. The SIACB \cite{barashev2012theoretical,golubov2014breakthrough} model assumes a value of 0.2. Our calculations indicate $\varepsilon^{g}$ is roughly independent of PKA energy (see \href{run:./SM/SM.tex}{SM - section 6}).

A sensitivity analysis of the SIACB model was performed by inputting the calculated $\varepsilon^{g}$ and $n^{g}$ values for M2R and M3R into our own implementation of the model \cite{barashev2015theoretical,golubov2014breakthrough}.
formulation of the model and its parameters are available in the \href{run:./SM/SM.tex}{SM - section 9}. Fig. \ref{fig:SIACB} illustrates these outcomes for prismatic strain. The choice of force-field has a material effect on $\varepsilon^{g}$ and $n^{g}$, but both estimates lead to a-type growth strains in broad agreement with experiments. However, negative strains along the prismatic directions are overestimated by the model, as compared to experimentally measured values (see \href{run:./SM/SM.tex}{SM - section 9}). Note rate-theory has a much larger effective timestep than MD. $\varepsilon^{g}$ would therefore also include some amount of annealing of cascade debris to more accurately estimate $\varepsilon^{g}$. As shown in the SM - section 7, $\varepsilon^{g}$ can increase as a function of annealing time. This suggests accelerated methods, such as object kinetic Monte Carlo (kMC) or on-the-fly kMC, can help get more accurate estimates of $\varepsilon^{g}$.

In conclusion, more likely than not, anisotropy of diffusion of cascade-induced SIA clusters is sufficient to explain RIG, but quantitative ambiguities remain given the sensitivity of our calculations to the choice of force-field. Also, \textit{a contrario} SIACB theory, we observed few perfect 1-D gliders and many 2-D gliders, which could materially contribute to growth.

Our simulations highlight the importance of developing high-fidelity force-fields with a focus on describing the energetics and kinetics of defects and defect clusters in Zr. Furthermore, our calculations suggest that the anisotropy of cascade-induced SIA clusters is greater than that of clusters formed by agglomeration of individual SIAs, which may in part explain why ion and neutron irradiation leads to greater RIG than electron irradiation.

% \section{Acknowledgments}
This work was supported by the Natural Sciences and Engineering Research Council of Canada (NSERC), and Atomic Energy of Canada Limited, under the auspices of the Federal Nuclear Science and Technology Program. The authors thank the Centre for Advanced Computing (CAC) at Queen's University and the Digital Research Alliance of Canada, formerly known as Compute Canada, for the generous allocation of computer resources.

% \clearpage
% \bibliographystyle{Setup/elsarticle-harv.bst}
\bibliographystyle{Setup/elsarticle-num.bst} 
\bibliography{Manuscript.bib}

\begin{thebibliography}{10}
\expandafter\ifx\csname url\endcsname\relax
  \def\url#1{\texttt{#1}}\fi
\expandafter\ifx\csname urlprefix\endcsname\relax\def\urlprefix{URL }\fi
\expandafter\ifx\csname href\endcsname\relax
  \def\href#1#2{#2} \def\path#1{#1}\fi

\bibitem{choi2013radiation}
S.~I. Choi, J.~H. Kim, Radiation-induced dislocation and growth behavior of zirconium and zirconium alloys--a review, Nuclear engineering and technology 45~(3) (2013) 385--392.

\bibitem{holt2008reactor}
R.~Holt, In-reactor deformation of cold-worked zr--2.5 nb pressure tubes, Journal of Nuclear Materials 372~(2-3) (2008) 182--214.

\bibitem{woo1998modeling}
C.~Woo, Modeling irradiation growth of zirconium and its alloys, Radiation effects and defects in solids 144~(1-4) (1998) 145--169.

\bibitem{christiaen2020influence}
B.~Christiaen, C.~Domain, L.~Thuinet, A.~Ambard, A.~Legris, Influence of vacancy diffusional anisotropy: Understanding the growth of zirconium alloys under irradiation and their microstructure evolution, Acta Materialia 195 (2020) 631--644.

\bibitem{holt1988mechanisms}
R.~Holt, Mechanisms of irradiation growth of alpha-zirconium alloys, Journal of Nuclear Materials 159 (1988) 310--338.

\bibitem{li2020cluster}
Y.~Li, N.~Ghoniem, Cluster dynamics modeling of irradiation growth in single crystal zr, Journal of Nuclear Materials 540 (2020) 152312.

\bibitem{christiaen2019new}
B.~Christiaen, C.~Domain, L.~Thuinet, A.~Ambard, A.~Legris, A new scenario for< c> vacancy loop formation in zirconium based on atomic-scale modeling, Acta Materialia 179 (2019) 93--106.

\bibitem{pugh1963properties}
S.~Pugh, Properties of reactor materials and the effects of radiation damage: Proceedings of the international conference held at berkeley castle, gloucestershire, england, 30 may--2 june 1961. dj littler (ed.), butterworths, london, 1962,{\pounds} 6 6s (1963).

\bibitem{woo1988theory}
C.~Woo, Theory of irradiation deformation in non-cubic metals: effects of anisotropic diffusion, Journal of Nuclear Materials 159 (1988) 237--256.

\bibitem{holt1993production}
R.~Holt, C.~Woo, C.~Chow, Production bias—a potential driving force for irradiation growth, Journal of nuclear materials 205 (1993) 293--300.

\bibitem{samolyuk2014analysis}
G.~D. Samolyuk, A.~V. Barashev, S.~I. Golubov, Y.~Osetsky, R.~E. Stoller, Analysis of the anisotropy of point defect diffusion in hcp zr, Acta materialia 78 (2014) 173--180.

\bibitem{barashev2015theoretical}
A.~V. Barashev, S.~I. Golubov, R.~E. Stoller, Theoretical investigation of microstructure evolution and deformation of zirconium under neutron irradiation, Journal of Nuclear Materials 461 (2015) 85--94.

\bibitem{golubov2014breakthrough}
S.~I. Golubov, A.~V. Barashev, R.~E. Stoller, B.~Singh, Breakthrough in understanding radiation growth of zirconiu, Tech. rep., Oak Ridge National Lab.(ORNL), Oak Ridge, TN (United States) (2014).

\bibitem{liu2020two}
S.-M. Liu, I.~J. Beyerlein, W.-Z. Han, Two-dimensional vacancy platelets as precursors for basal dislocation loops in hexagonal zirconium, Nature communications 11~(1) (2020) 1--8.

\bibitem{dai2019mechanism}
C.~Dai, P.~Saidi, M.~Topping, L.~B{\'e}land, Z.~Yao, M.~Daymond, A mechanism for basal vacancy loop formation in zirconium, Scripta Materialia 172 (2019) 72--76.

\bibitem{dai2021stability}
C.~Dai, C.~Varvenne, P.~Saidi, Z.~Yao, M.~R. Daymond, L.~K. B{\'e}land, Stability of vacancy and interstitial dislocation loops in $\alpha$-zirconium: Atomistic calculations and continuum modelling, Journal of Nuclear Materials 554 (2021) 153059.

\bibitem{harte2017effect}
A.~Harte, D.~J{\"a}dern{\"a}s, M.~Topping, P.~Frankel, C.~Race, J.~Romero, L.~Hallstadius, E.~C. Darby, M.~Preuss, The effect of matrix chemistry on dislocation evolution in an irradiated zr alloy, Acta Materialia 130 (2017) 69--82.

\bibitem{dai2018atomistic}
C.~Dai, Atomistic simulations of irradiation-induced dislocation loops in zirconium alloys, Ph.D. thesis (2018).

\bibitem{voskoboinikov2006atomic}
R.~E. Voskoboinikov, Y.~N. Osetsky, D.~J. Bacon, Atomic-scale simulation of defect cluster formation in high-energy displacement cascades in zirconium (2006).

\bibitem{gao2001temperature}
F.~Gao, D.~Bacon, L.~Howe, C.~So, Temperature-dependence of defect creation and clustering by displacement cascades in $\alpha$-zirconium, Journal of nuclear materials 294~(3) (2001) 288--298.

\bibitem{voskoboinikov2006identification}
R.~Voskoboinikov, Y.~N. Osetsky, D.~Bacon, Identification and morphology of point defect clusters created in displacement cascades in $\alpha$-zirconium, Nuclear Instruments and Methods in Physics Research Section B: Beam Interactions with Materials and Atoms 242~(1-2) (2006) 530--533.

\bibitem{whiting1996computer}
B.~Whiting, D.~Bacon, Computer simulation of the migration of self-interstitial atoms in alpha-zirconium, MRS online proceedings library 439~(1) (1996) 389--394.

\bibitem{de2002mobility}
N.~De~Diego, Y.~N. Osetsky, D.~Bacon, Mobility of interstitial clusters in alpha-zirconium, Metallurgical and materials transactions A 33~(13) (2002) 783--789.

\bibitem{de2011structure}
N.~De~Diego, A.~Serra, D.~Bacon, Y.~N. Osetsky, On the structure and mobility of point defect clusters in alpha-zirconium: a comparison for two interatomic potential models, Modelling and simulation in materials science and engineering 19~(3) (2011) 035003.

\bibitem{march2022defect}
J.~F. March-Rico, B.~D. Wirth, Defect cluster configurations and mobilities in $\alpha$-zr: A comparison of the bmd19 and m07 interatomic potentials, Journal of Nuclear Materials 559 (2022) 153441.

\bibitem{maxwell2020atomistic}
C.~Maxwell, J.~Pencer, E.~Torres, Atomistic simulation study of clustering and evolution of irradiation-induced defects in zirconium, Journal of Nuclear Materials 531 (2020) 151979.

\bibitem{stoller2016impact}
R.~Stoller, A.~Tamm, L.~B{\'e}land, G.~Samolyuk, G.~Stocks, A.~Caro, L.~Slipchenko, Y.~N. Osetsky, A.~Aabloo, M.~Klintenberg, et~al., Impact of short-range forces on defect production from high-energy collisions, Journal of chemical theory and computation 12~(6) (2016) 2871--2879.

\bibitem{beland2016atomistic}
L.~K. B{\'e}land, Y.~N. Osetsky, R.~E. Stoller, Atomistic material behavior at extreme pressures, npj Computational Materials 2~(1) (2016) 1--4.

\bibitem{beland2017accurate}
L.~K. B{\'e}land, A.~Tamm, S.~Mu, G.~D. Samolyuk, Y.~N. Osetsky, A.~Aabloo, M.~Klintenberg, A.~Caro, R.~E. Stoller, Accurate classical short-range forces for the study of collision cascades in fe--ni--cr, Computer Physics Communications 219 (2017) 11--19.

\bibitem{byggmastar2018effects}
J.~Byggm{\"a}star, F.~Granberg, K.~Nordlund, Effects of the short-range repulsive potential on cascade damage in iron, Journal of Nuclear Materials 508 (2018) 530--539.

\bibitem{becquart2021modelling}
C.~S. Becquart, A.~De~Backer, P.~Olsson, C.~Domain, Modelling the primary damage in fe and w: Influence of the short range interactions on the cascade properties: Part 1--energy transfer, Journal of Nuclear Materials 547 (2021) 152816.

\bibitem{mendelev2007development}
M.~I. Mendelev, G.~J. Ackland, Development of an interatomic potential for the simulation of phase transformations in zirconium, Philosophical Magazine Letters 87~(5) (2007) 349--359.

\bibitem{LAMMPS}
A.~P. Thompson, H.~M. Aktulga, R.~Berger, D.~S. Bolintineanu, W.~M. Brown, P.~S. Crozier, P.~J. in~'t Veld, A.~Kohlmeyer, S.~G. Moore, T.~D. Nguyen, R.~Shan, M.~J. Stevens, J.~Tranchida, C.~Trott, S.~J. Plimpton, {LAMMPS} - a flexible simulation tool for particle-based materials modeling at the atomic, meso, and continuum scales, Comp. Phys. Comm. 271 (2022) 108171.
\newblock \href {https://doi.org/10.1016/j.cpc.2021.108171} {\path{doi:10.1016/j.cpc.2021.108171}}.

\bibitem{saidi2020method}
P.~Saidi, Z.~Wang, Y.~Mao, L.~K. B{\'e}land, Z.~Yao, M.~R. Daymond, A method for calculation of bias factor in anisotropic mediums, application to $\alpha$- zirconium, Journal of Nuclear Materials 528 (2020) 151882.

\bibitem{ovito}
A.~Stukowski, {Visualization and analysis of atomistic simulation data with OVITO-the Open Visualization Tool}, {MODELLING AND SIMULATION IN MATERIALS SCIENCE AND ENGINEERING} {18}~({1}) ({JAN} {2010}).
\newblock \href {https://doi.org/{10.1088/0965-0393/18/1/015012}} {\path{doi:{10.1088/0965-0393/18/1/015012}}}.

\bibitem{nordlund2018improving}
K.~Nordlund, S.~J. Zinkle, A.~E. Sand, F.~Granberg, R.~S. Averback, R.~Stoller, T.~Suzudo, L.~Malerba, F.~Banhart, W.~J. Weber, et~al., Improving atomic displacement and replacement calculations with physically realistic damage models, Nature communications 9~(1) (2018) 1--8.

\bibitem{yang2021full}
Q.~Yang, P.~Olsson, Full energy range primary radiation damage model, Physical Review Materials 5~(7) (2021) 073602.

\bibitem{mendelev2012development}
M.~Mendelev, M.~Kramer, S.~Hao, K.~Ho, C.~Wang, Development of interatomic potentials appropriate for simulation of liquid and glass properties of nizr2 alloy, Philosophical Magazine 92~(35) (2012) 4454--4469.

\bibitem{wilson2015anisotropy}
S.~Wilson, M.~Mendelev, Anisotropy of the solid--liquid interface properties of the ni--zr b33 phase from molecular dynamics simulation, Philosophical Magazine 95~(2) (2015) 224--241.

\bibitem{yang2018molecular}
X.~Yang, X.~Zeng, L.~Chen, Y.~Guo, H.~Chen, F.~Wang, Molecular dynamics simulations of the primary irradiation damage in zirconium, Nuclear Instruments and Methods in Physics Research Section B: Beam Interactions with Materials and Atoms 436 (2018) 92--98.

\bibitem{terentyev2006effect}
D.~Terentyev, C.~Lagerstedt, P.~Olsson, K.~Nordlund, J.~Wallenius, C.~S. Becquart, L.~Malerba, Effect of the interatomic potential on the features of displacement cascades in $\alpha$-fe: A molecular dynamics study, Journal of nuclear materials 351~(1-3) (2006) 65--77.

\bibitem{calder2010origin}
A.~Calder, D.~J. Bacon, A.~V. Barashev, Y.~N. Osetsky, On the origin of large interstitial clusters in displacement cascades, Philosophical Magazine 90~(7-8) (2010) 863--884.

\bibitem{beland2013replenish}
L.~K. B{\'e}land, Y.~Anahory, D.~Smeets, M.~Guihard, P.~Brommer, J.-F. Joly, J.-C. Pothier, L.~J. Lewis, N.~Mousseau, F.~Schiettekatte, Replenish and relax: Explaining logarithmic annealing in ion-implanted c-si, Physical review letters 111~(10) (2013) 105502.

\bibitem{beland2013long}
L.~K. B{\'e}land, N.~Mousseau, Long-time relaxation of ion-bombarded silicon studied with the kinetic activation-relaxation technique: Microscopic description of slow aging in a disordered system, Physical Review B 88~(21) (2013) 214201.

\bibitem{beland2015kinetic}
L.~K. B{\'e}land, Y.~N. Osetsky, R.~E. Stoller, H.~Xu, Kinetic activation--relaxation technique and self-evolving atomistic kinetic monte carlo: Comparison of on-the-fly kinetic monte carlo algorithms, Computational Materials Science 100 (2015) 124--134.

\bibitem{brommer2014understanding}
P.~Brommer, L.~K. B{\'e}land, J.-F. Joly, N.~Mousseau, Understanding long-time vacancy aggregation in iron: A kinetic activation-relaxation technique study, Physical Review B 90~(13) (2014) 134109.

\bibitem{beland2015slow}
L.~K. B{\'e}land, Y.~N. Osetsky, R.~E. Stoller, H.~Xu, Slow relaxation of cascade-induced defects in fe, Physical Review B 91~(5) (2015) 054108.

\bibitem{beland2016differences}
L.~K. B{\'e}land, G.~D. Samolyuk, R.~E. Stoller, Differences in the accumulation of ion-beam damage in ni and nife explained by atomistic simulations, Journal of Alloys and Compounds 662 (2016) 415--420.

\bibitem{barashev2001reaction}
A.~Barashev, S.~Golubov, H.~Trinkaus, Reaction kinetics of glissile interstitial clusters in a crystal containing voids and dislocations, Philosophical Magazine A 81~(10) (2001) 2515--2532.

\bibitem{carpenter1988irradiation}
G.~Carpenter, R.~Zee, A.~Rogerson, Irradiation growth of zirconium single crystals: A review, Journal of Nuclear Materials 159 (1988) 86--100.

\bibitem{NORGETT197550}
M.~Norgett, M.~Robinson, I.~Torrens, \href{https://www.sciencedirect.com/science/article/pii/0029549375900357}{A proposed method of calculating displacement dose rates}, Nuclear Engineering and Design 33~(1) (1975) 50--54.
\newblock \href {https://doi.org/https://doi.org/10.1016/0029-5493(75)90035-7} {\path{doi:https://doi.org/10.1016/0029-5493(75)90035-7}}.
\newline\urlprefix\url{https://www.sciencedirect.com/science/article/pii/0029549375900357}

\bibitem{ackland1995defect}
G.~Ackland, S.~Wooding, D.~Bacon, Defect, surface and displacement-threshold properties of $\alpha$-zirconium simulated with a many-body potential, Philosophical Magazine A 71~(3) (1995) 553--565.

\bibitem{barashev2012theoretical}
A.~V. Barashev, S.~I. Golubov, R.~E. Stoller, Theoretical investigation of microstructure evolution and deformation of zirconium under cascade damage conditions, ORNL/TM-2012 225 (2012).

\end{thebibliography}
% \bibliographystyle{unsrt}
% \bibliography{xampl}

\end{document}

\begin{frontmatter}
\title{Anisotropic diffusion of radiation-induced self-interstitial clusters in hcp zirconium: a molecular dynamics and rate-theory assessment: Supplementary Materials} 

%%%%%%AUTHORS%%%%%%%%%%
\author[1]{Amir Ghorbani}
\author[1]{Yu Luo}
\author[1,2]{Peyman Saidi}
\author[1]{Laurent Karim Béland\corref{cor1}}

\ead{laurent.beland@queensu.ca}
\cortext[cor1]{Corresponding author}

\address[1]{Department of Mechanical and Materials Engineering, Queen's University, Kingston, ON K7L3N6, Canada}
\address[2]{Canadian Nuclear Laboratories, Chalk River, ON K0J 1J0, Canada}
\date{June 2023}

% \section{Abstract}
\begin{abstract}
In this supplementary materials, the reparameterization of Mendelev and Ackland 2 \& 3 (M2 and M3) embedded atom model (EAM) potentials, is explained in detail. A comparison of the re-stiffened potentials (M2R and M3R) with density functional theory benchmarks is presented. Next, the procedure for measuring the diffusion of self-interstitial atoms (SIA) clusters produced by cascades and agglomeration of individual SIAs is explained. Furthermore, we show how the ratio of a-biased diffusing SIA clusters to total SIAs, also known as epsilon ($\varepsilon_{i}^{g}$), varies according to choices made during the analysis. Finally, we provide more details pertaining to the SIA cluster bias (SIACB) rate-theory model.
\end{abstract}

\begin{keyword}
Irradiation \sep Radiation Induced Growth  \sep Cluster \sep Zirconium \sep Molecular Dynamics
% \sep Nuclear Materials
\end{keyword}
\end{frontmatter}

%%%%%%%%%%%%%%%%%%%%%MAIN TEXT%%%%%%%%%%%%%%%%%%%%%%%%%%
% \section{Main Body}
% \begin{multicols}{2}
\section{Computational Packages}
All density functional theory (DFT) calculations in this research were performed using Quantum Espresso \cite{giannozzi2009quantum}. All Embedded atom model (EAM) simulations were performed using the Large-scale Atomic/Molecular Massively Parallel Simulator (LAMMPS) \cite{LAMMPS}. Analysis and figures were created using Python (matplotlib) and packages such as OVITO \cite{stukowski2009visualization}, Atomic Simulation Environment (ASE) \cite{larsen2017atomic}, and atsim.potentials \cite{atsim2020}. 

\section{Reparameterization of interatomic potential}
Force-fields used to simulate collision cascades should be able to accurately model situations involving close atomic distances. The choice of potential can have an important effect on the 1-D/3D movement of clusters [2]. Hence, we aimed to reparametrize two EAM potentials by Mendelev and Ackland \cite{mendelev2007development} (referred to as M2 and M3) at short interatomic distances, while minimizing the impact of the reparameterization on near-equilibrium physical properties of the material.

\subsection{Comparison of M2 and M3 Potentials}
Mendelev and Ackland \cite{mendelev2007development} prepared M2 to model the hcp-bcc and liquid-solid phase transitions in Zr. They also created M3 potential which describes radiation-damage in zr-hcp better \cite{tian2021radiation}. A good number of EAM modelling attempts of Zr have been based on M2 and M3 interatomic potentials, e.g. \cite{torres2023effect,barik2022molecular,zhang2022crack,bhatia2013energetics}. Also, they are compared in different aspects \cite{nicholls2023transferability}. M3 has a more realistic vacancy migration energy, but M2 has a better vacancy binding energy \cite{varvenne2014vacancy}. They both exhibit similar SIA formation energies and migration energies. Dai \textit{et al.} \cite{dai2019deformation} showed that M3 offers a better description of prism vs basal stacking fault energy, which leads to different SIA cluster behavior. In the same study, nanotwins were observed with M3, but not M2. In general, both M2 and M3 offer reasonable descriptions of generalized stacking fault energies and elastic constants; M3 stacking fault energy values are in closer agreement with DFT \cite{varvenne2014vacancy}. According to ab initio MD \cite{yang2021full}, the threshold displacement energy (TDE) in the [0001] direction is 20 eV. Both M2 and M3 overestimate this value, but M2 is closer. It is difficult to make a case that one of these two interatomic potentials is truly better than the other. They possess complementary strengths and limitations for studying point defects and defect clusters. Using them both was helpful in identifying which features of our simulations are likely to be generalizable and which features are potential-dependent – and therefore maybe not transposable. 

\subsection{Method}
\begin{figure}[!t]
\centering
\includegraphics[width=0.6\linewidth]{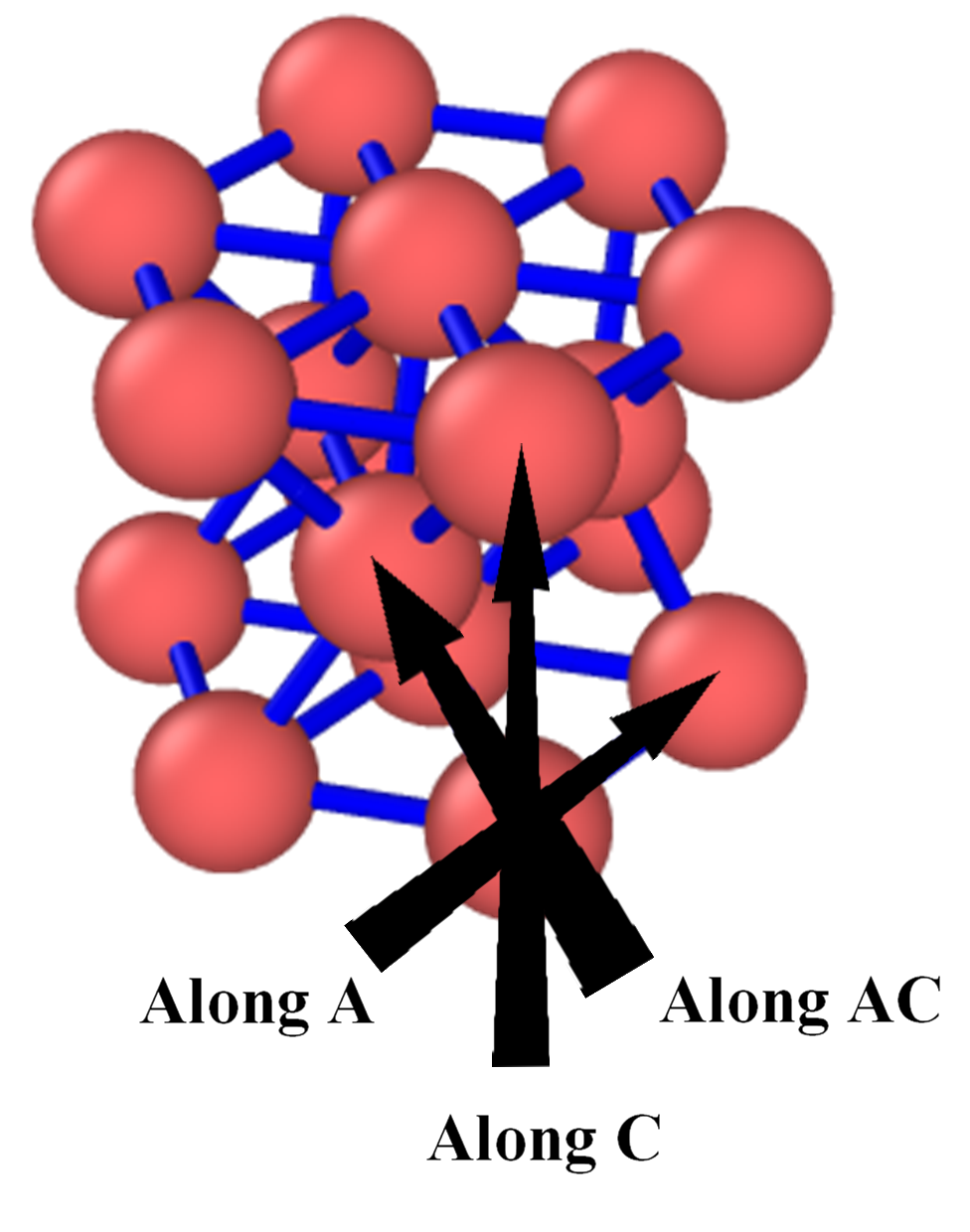}
\caption{Schematic representation of embedded dimer along A, C and AC directions.}
\label{fig:EmbeddedDimerSchematic}
\end{figure}

\begin{figure}[!t]
\centering
\includegraphics[width=0.65\linewidth]{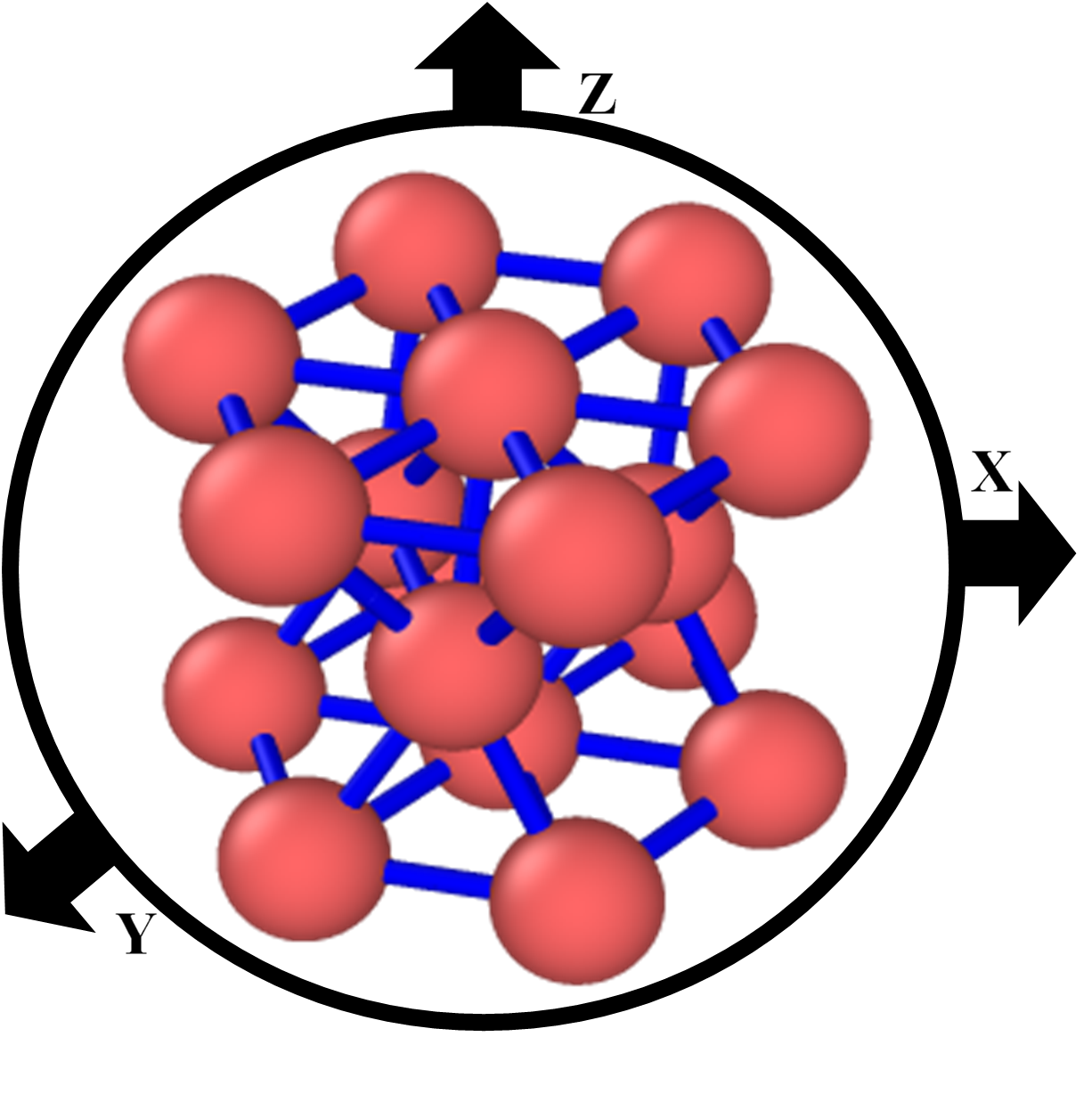}
\caption{Schematic representation of EOS}
\label{fig:EquationOfStateSchematic}
\end{figure}

The EAM pair term was reparametrized using embedded dimer DFT calculations, also known as quasi-static drag. This refers to taking an atom in a perfect crystal and dragging it towards a neighbouring atom.  Three embedded dimers were considered, as illustrated in Fig \ref{fig:EmbeddedDimerSchematic}: toward the nearest atom in the same plane (A), toward the nearest atom in the concurrent plane (C), diagonally toward the nearest atom in the next plane (AC). To reparametrize the embedding term, we performed DFT calculations involving a perfect hcp crystal upon which uniform compression and tension were applied, i.e. we calculated its equation of state (EOS). The EOS calculations are illustrated in Fig. \ref{fig:EquationOfStateSchematic}. \Cref{table:Dft} shows the details of these DFT calculations. Here, the generalized gradient approximation exchange-correlation (XC) functional by Perdew-Burke-Ernzerhof (PBE) was chosen. Ion/cell dynamics was performed using the Broyden–Fletcher–Goldfarb–Shanno (BFGS) algorithm. Furthermore, the pseudopotential file named zr-pbe-v1.uspp.F.UPF (GBRV) was used \cite{garrity2014pseudopotentials}.

\begin{table}[t]
\centering
\caption{Details of DFT calculations}
\begin{tabular}{ll}
Parameter                       & Value             \\ \hline
Calculation                     & scf, relax        \\
no. of optimization steps       & 100               \\
no. of atoms                    & 32                \\
XC functional                   & PBE               \\
smearing                        & Methfessel-Paxton \\
max. iterations in a scf        & 1000              \\
ionic and cell dynamics         & bfgs              \\
k-point grid                    & (6 4 4)           \\
wavefunctions cutoff (Ry)       & 50               
\label{table:Dft}
\end{tabular}
\end{table}

\begin{figure}[!t]
\centering
\includegraphics[width=0.82\linewidth]{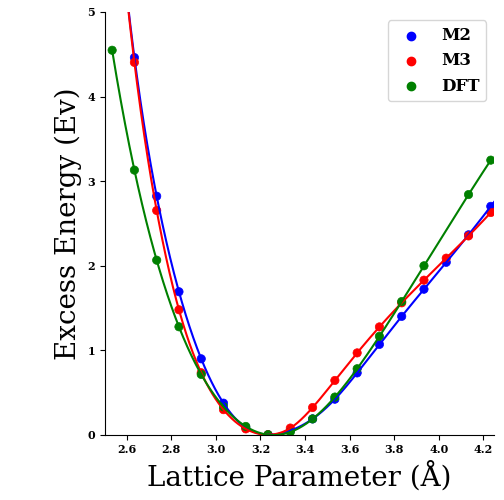}
\caption{EOS calculations using the original M2 and M3 potentials in comparison with DFT results. }
\label{fig:EquationOfState}
\end{figure}

\begin{figure*}
\centering
\includegraphics[width=0.75\paperwidth]{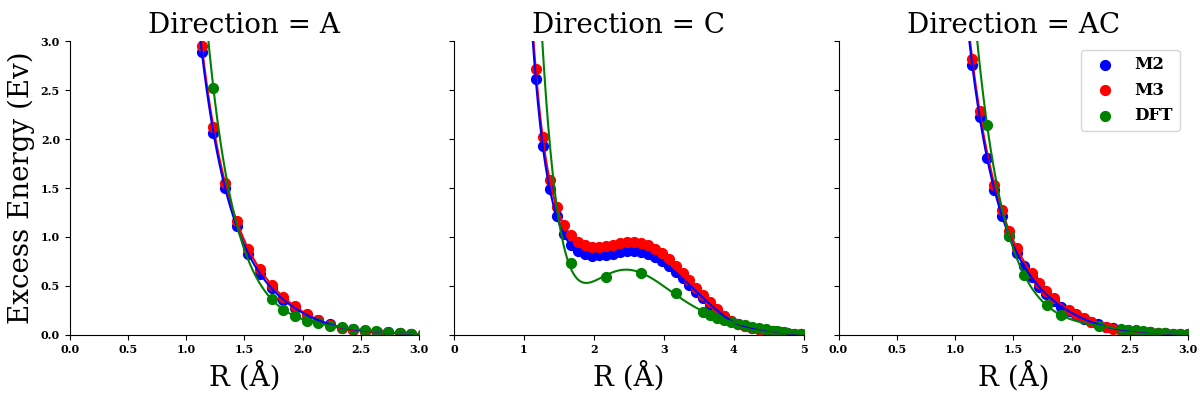}
\caption{ Three embedded dimers (also known as quasi-static drag) were considered, with displacement along A, C and AC directions. This figure compares the excess energy as calculated with M2 and M3 with DFT calculations. Inconsistencies are minor except in the C direction when the embedded atom passes through the adjacent basal plane.}
\label{fig:EdOriginal}
\end{figure*}

\Cref{fig:EquationOfState} shows the measured excessive energies for M2 and M3 in EOS as compared to DFT results. While M2 and M3 overestimate the energy in contraction, the divergence is greatest in expansion. Also, as is seen in \cref{fig:EdOriginal}, there are minor inconsistencies between DFT and the original EAM calculations of embedded dimer in A, C and AC directions. The divergence between EAM and DFT is greatest for the C direction (e.g. 0.3 eV at 2.5Å). These small deviations suggest the original M2 and M3 potentials would likely have led to reasonable primary damage production predictions, even without reparameterization. Nonetheless, given this validation required performing the DFT calculations, we opted to reparametrize the force-fields using these DFT calculations, as it provides a small increment in accuracy at little cost. 

\begin{figure*}
\centering
\includegraphics[width=0.75\paperwidth]{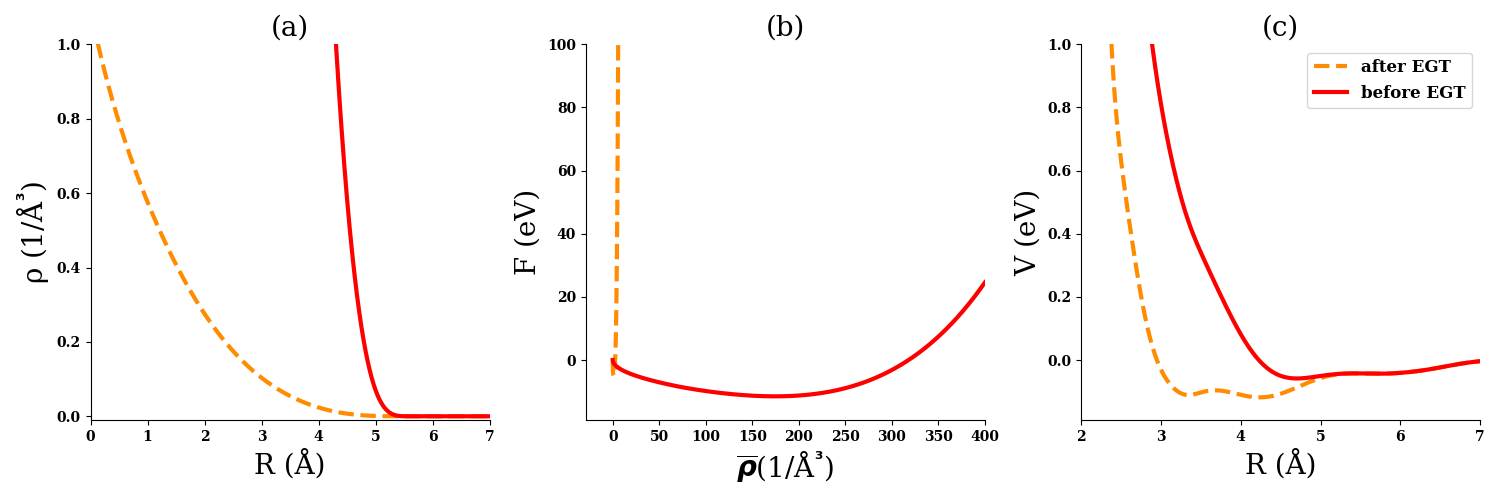}
\caption{The M3 potential before and after effective gauge transformation. (a) Electron density, (b) embedded part, (c) pair part. }
\label{fig:Egt}
\end{figure*}

To reparametrize the potential, we first applied the effective gauge transformation (EGT), as described by Pasianot \cite{bonny2010gauge}. The electron density ($\rho$) is re-scaled so that it is equal to unity at the equilibrium lattice parameter using \cref{G1}. Next, energy is exchanged between the pair part ($V$) and the embedded function ($F$) using \cref{G2} to ensure the derivative of $F$ is zero at the equilibrium lattice parameter. S and C constants are chosen as the inverse value of electron density  and derivative of  as shown by \cref{SConstant} and \cref{CConstant}, respectively, at equilibrium lattice distance of a perfect Zr-hcp crystal structure (3.23 Å).
\Cref{fig:Egt} show the electron density (a), embedded part (b) and pair part (c) curves of M3 before and after EGT. Here, energies are in units of eV, the electron density is in Å$^{-3}$, and the radius ($R$) is in Å \cite{foiles1986embedded}. The EAM model is invariant to any transformation in which a term linear in the electron density is added to or subtracted from the embedding function and an appropriate adjustment is made to the potential \cite{johnson1989analytic}. In other words, the EGT changes the functional form without modifying its energetic description \cite{bonny2010gauge}, and minimizes the nonlinear contribution of the embedding term \cite{caro2005classical}. In our experience, this representation facilitates reparameterization of the potential at a short distance.

\begin{equation}
  G_{1} =
    \begin{cases}
      \widetilde{\rho}_{(r)} = S \rho_{(r)} \\
      \widetilde{F}_{(\widetilde{\bar{\rho}})} = \widetilde{F}_{(\widetilde{\bar{\rho}}/S)}
    \end{cases}
    \label{G1}
\end{equation}

\begin{equation}
  G_{2} =
    \begin{cases}
      \widetilde{V}_{(r)} = V_{(r)} + 2C\rho_{(r)} \\
      \widetilde{F}_{(\bar{\rho})} = \widetilde{F}_{(\bar{\rho})} - C\bar{\rho}
    \end{cases}
    \label{G2}
\end{equation}

\begin{equation}
S = 1/\bar{\rho}_{(r_{eq})}
\label{SConstant}
\end{equation}

\begin{equation}
C=\acute{F}_{\bar\rho_{eq}}
\label{CConstant}
\end{equation}

\begin{figure}[H]
    \begin{tikzpicture}
    \node(a){\includegraphics[width=0.75\linewidth]{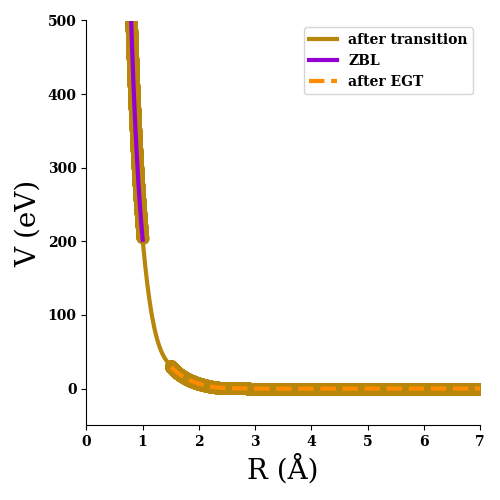}};
    \node at (a.north east)
    [
    anchor=center,
    xshift=-30mm,
    yshift=-30mm
    ]
    {
    };
    \end{tikzpicture}
\vspace{-2em}
\caption{Spline between the pair part of M3 (after EGT) and ZBL potentials.}
\label{fig:ZblSpline}
\end{figure}
\vspace{-1em}

After applying the effective gauge transformation, electron density was modified so to be saturated at ultra-close interatomic distances: the electron density function was considered to remain constant for interatomic distances of less than 1Å. Furthermore, the smooth transition of the first derivative of the electron density function necessitated a transition region of 1±0.2 Å.

\begin{figure*}
\centering
\includegraphics[width=0.75\paperwidth]{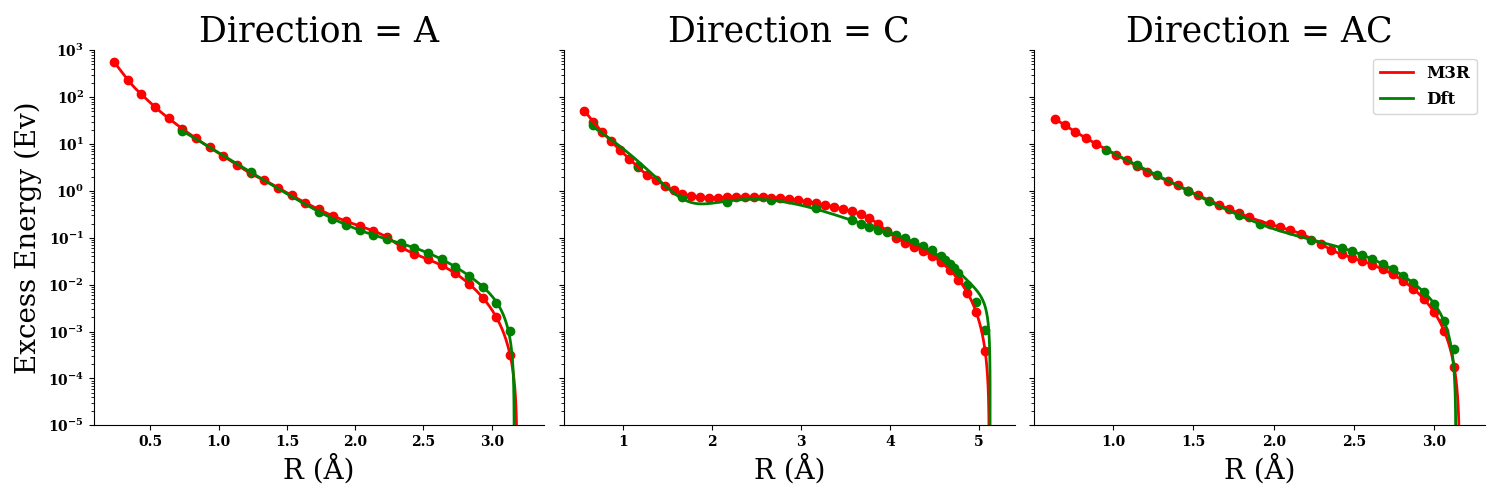}
\caption{The effect of reparameterization of M3 on excess energy of an embedded dimer along A, C and AC directions. Reparameterized M3 (M3R) excess energy is closer to DFT results.}
\label{fig:M3rEmbeddedDimer}
\end{figure*}

Third, the ZBL potential was connected to the EAM pair term using cubic splines over a transition region starting from 1Å and finishing at 1.25Å. These spline cutoff distances led to a good agreement with our embedded dimer DFT calculations, with very little effect on near-equilibrium energetics. \Cref{fig:ZblSpline} shows the smooth transition between ZBL and the pair part of M3.

Finally, the deviations between EAM-based and DFT-based EOS were minimized by modifying the embedded part, which resulted in reparametrizated potentials (M2R and M3R). We chose the cost function displayed in \cref{CostFunction} to minimize the deviations \cite{stoller2016impact}. \Cref{fig:M3rEmbeddedDimer} shows the embedded dimer calculations along A, C and AC directions for M3 after reparameterization (M3R) next to DFT results used as a benchmark. As is seen, the optimization shifted the EAM energy values towards DFT values and lessened the divergence, especially along the C direction (e.g. 0.07 eV at 2.5Å). Furthermore, the reparameterized potentials (M2R and M3R) can now observe short atomic distances compared to the original potentials (M2 and M3). It is worth mentioning that an overshoot in a fraction of the transition region in the AC direction is observed.

% \vspace{-0.75em}
\begin{equation}
CF=\sum_{i=1}^N\sqrt{\mid E_{DFT}(r_{i})-E_{EAM}(r_{i})}
\label{CostFunction}
\end{equation}

\subsection{Assessment}
\begin{figure}[H]
\centering
\includegraphics[width=0.65\linewidth]{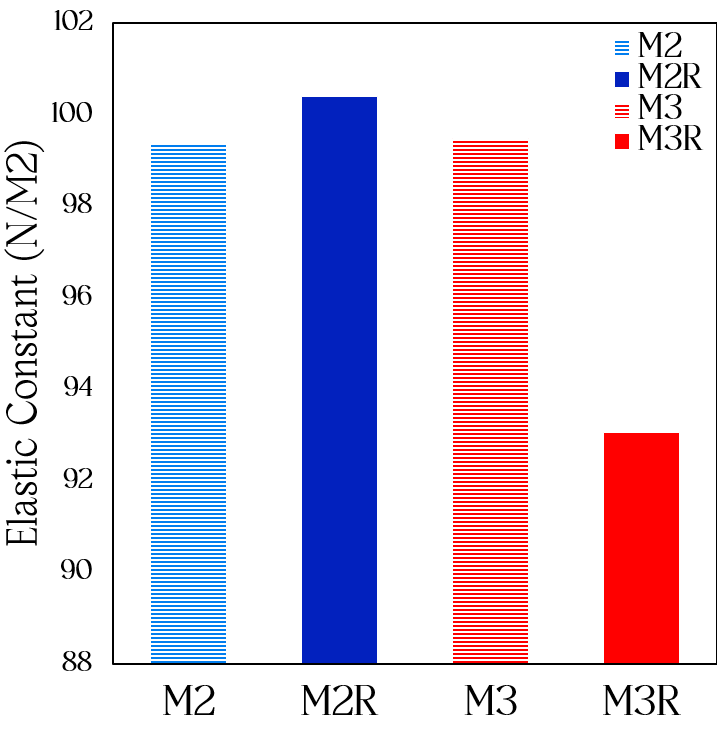}
\vspace{\MyFigSpace}
\caption{Bulk modulus.}
\label{fig:BulkModulus}
\end{figure}

\begin{figure}[H]
\centering
\includegraphics[width=0.65\linewidth]{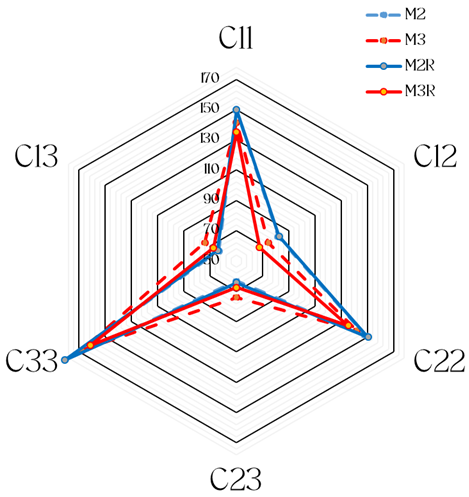}
\vspace{\MyFigSpace}
\caption{Components of the elastic constant tensor.}
\label{fig:ComponentsOfElasticConstantTensor}
\end{figure}

\begin{figure}[H]
\centering
\includegraphics[width=0.65\linewidth]{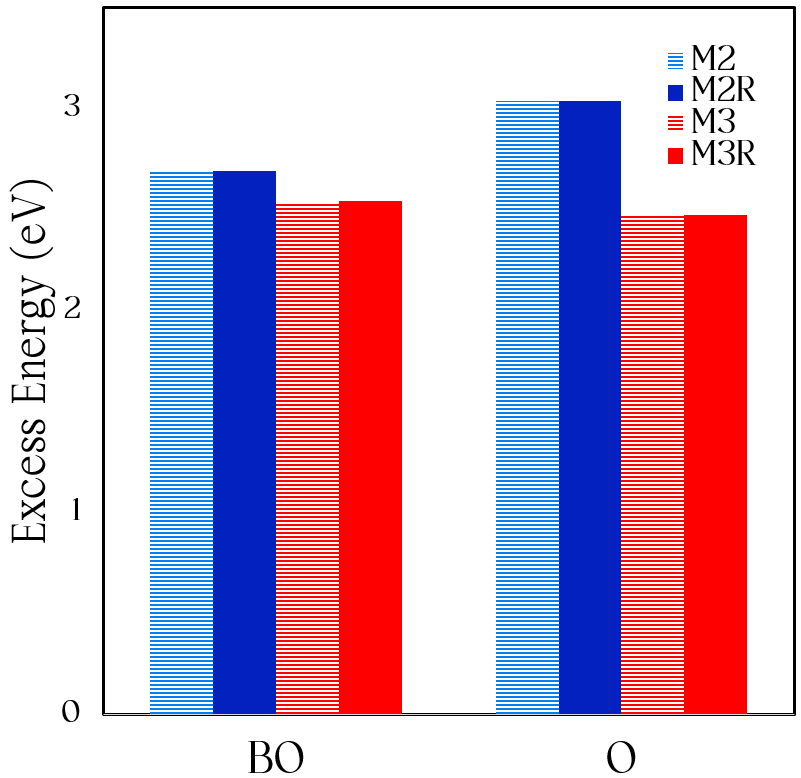}
\vspace{\MyFigSpace}
\caption{SIA Formation energy}
\label{fig:SiaFormation}
\end{figure}

Changes in the bulk modulus and the components of the elastic constant tensor are reported in \cref{fig:BulkModulus} and \cref{fig:ComponentsOfElasticConstantTensor}. The effects of reparameterization of M2 on its near-equilibrium properties are negligible; reparameterization of M3 led to a \%6 decrease in elastic constants. Basal-octahedral (BO) and octahedral (O) self-interstitial atoms (SIA) were introduced in a perfect structure which are the first and fifth stable SIA forms in hcp structure, respectively \cite{peng2014pressure}. The comparison shows that in each case, minimization causes less than 0.1 percent increase in formation energy, which is illustrated in \cref{fig:SiaFormation}.

We verified if the TDE was modified due to the reparameterization. We performed these simulations in the NVE ensemble using LAMMPS, after equilibrating a 7128-atom Zr hcp crystal at 10 Kelvin. The TDE was estimated by gradually increasing the PKA energy from 0 eV to 200 eV with a step of 2 eV. Four primary knock-on atom directions were considered. As illustrated in \cref{table:Tde}, our reparameterization had a minor effect on TDEs, within the resolution of our calculations.

\section{Agglomeration simulation}
\subsection{Setup}
\begin{figure}[H]
\centering
\includegraphics[width=0.75\linewidth]{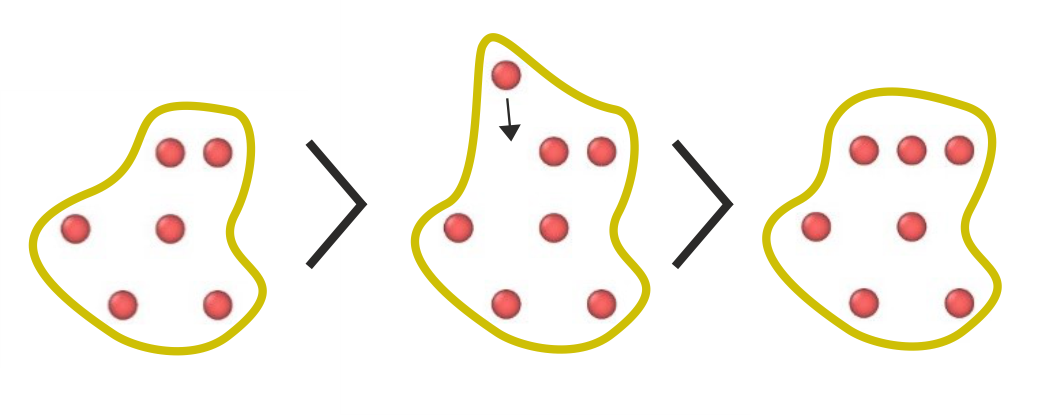}
\vspace{\MyFigSpace}
\caption{Schematic representation of Agglomeration method.}
\label{fig:Agglomeration}
\end{figure}

The simulation box in this setup is comprised of 15,000 atoms under periodic boundary conditions (PBC). The clusters are equilibrated at 573K and 0 Pa. Then, the clusters’ diffusion coefficients are calculated in the microcanonical ensemble. To agglomerate clusters, SIAs were added in a random location near the existing clusters one at a time. Next, enough time was given for the SIAs to form a single cluster (in the NPT ensemble), followed by an NVE run to accumulate diffusion statistics. We considered clusters of up to 20 SIAs. \Cref{fig:Agglomeration} illustrates atomic configurations as the SIAs form a cluster. In this figure, a single SIA is spawned near a cluster which has 6 SIAs to form a 7 SIA cluster.  \Cref{table:AggregationUnitCell} defines three primitive vectors of unit cell and four basis atoms at 0K. \Cref{table:AggregationStages} shows the details of agglomeration stages.

\vspace{-0.75em}
\begin{table}[H]
\centering
\caption{Threshold displacement energies in primary directions for M2, M2R, M3 and M3R.}
\vspace{\MyFigSpace}
\begin{tabular}{c|cccc}
& {[}0001{]} & {[}$\bar{1}$2$\bar{1}$0{]} & {[}$\bar{1}$100{]} & {[}01$\bar{1}$0{]} \\ \hline
M2 & 34 & 20 & 18 & 22 \\
M2R & 32 & 20 & 18 & 22 \\
M3 & 44 & 32 & 18 & 18 \\
M3R & 46 & 30 & 18 & 16
\label{table:Tde}
\end{tabular}
\end{table}

\vspace{\MyFigSpace}

\vspace{\MyFigSpace}

\vspace{\MyFigSpace}

\vspace{\MyFigSpace}

\begin{table}[H]
\caption{Unit cell vector and basis.}
\vspace{\MyFigSpace}
\centering
\begin{tabular}{ll}
Parameter                    & Value                               \\ \hline
\multicolumn{1}{l|}{A1}      & \textless{}1,0,0\textgreater{}      \\
\multicolumn{1}{l|}{A2}      & \textless{}0,1.7324,0\textgreater{} \\
\multicolumn{1}{l|}{A3}      & \textless{}0,0,1.6\textgreater{}    \\
\multicolumn{1}{l|}{Basis 1} & {[}0,0,0{]}                         \\
\multicolumn{1}{l|}{Basis 2} & {[}0.5,0.5,0{]}                     \\
\multicolumn{1}{l|}{Basis 3} & {[}0.5, 0.1667,0.5{]}               \\
\multicolumn{1}{l|}{Basis 4} & {[}0, 0.6667, 0.5{]}               
\end{tabular}
\label{table:AggregationUnitCell}
\end{table}

\begin{table*}[!t]
\caption{LAMMPS setup for simulation of clusters produced by aggregation of SIAs.}
\vspace{\MyFigSpace}
\centering
\begin{tabular}{llllll}
\multicolumn{2}{l}{Stage}                   & Ensemble & Steps       & Dump interval & Max. timestep (ps) \\ \hline
1. & \multicolumn{1}{l|}{Heating to 573K}   & NPT      & 500,000     & 10000         & 0.002         \\
2. & \multicolumn{1}{l|}{Relaxation}        & NVE      & 100,000     & 10000         & 0.002         \\
3. & \multicolumn{1}{l|}{Addition}          & NVE      & 1           & 1             & 0.0000001     \\
4. & \multicolumn{1}{l|}{Clustering (573K)} & NPT      & 100,000     & 500           & 0.0001        \\
5. & \multicolumn{1}{l|}{Aging}             & NVE      & 100,000,000 & 500000        & 0.001        
\end{tabular}
\label{table:AggregationStages}
\end{table*}

% \vspace{-3em}

\subsection{Mean square displacement}
\begin{figure*}[!t]
\centering
\includegraphics[width=0.80\paperwidth]{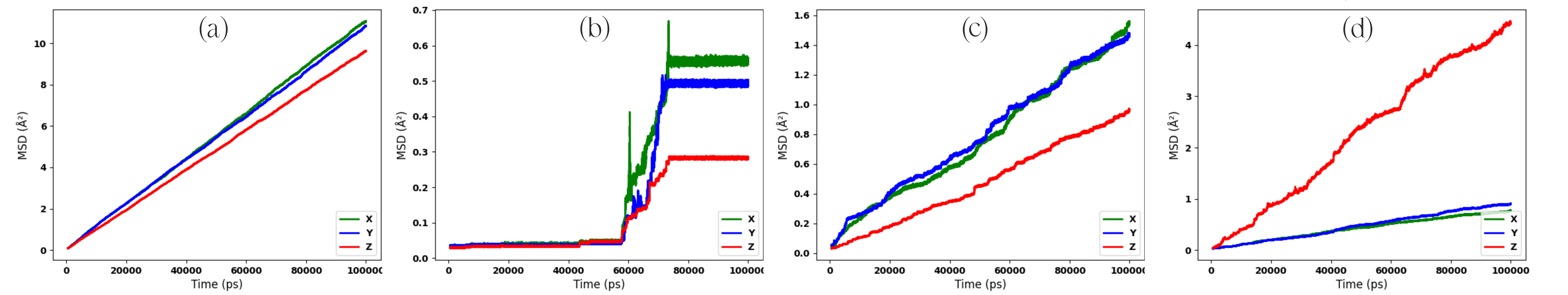}
\vspace{\MyFigSpace}
\caption{Accumulative MSD measured against time at 573K for (a) single SIA, (b) SIA cluster of size 6, (c) SIA cluster of size 7, (d) SIA cluster of size 14}
\label{fig:MsdSynthetic}
\end{figure*}

A fix in LAMMPS was used to measure the mean square displacement (MSD) of atoms. Since only one cluster is present in each run, the MSD of atoms represents the movement of clusters and was used to trace them. The slope of this MSD curve against time was calculated using Python which later is used to determine the diffusion coefficients. \Cref{fig:MsdSynthetic} shows these curves for a single SIA and clusters of different sizes. Occasionally, clusters fall into semi-stable configurations and stop gliding. This trap is characterized by plateau lines in MSD curves as seen for the 6-SIA cluster which does not construct the diffusion curve. This behaviour is reproducible. We only considered the non-trapped diffusion regime for our MSD calculations, since the goal was to measure the anisotropy of diffusion. Adding one SIA to a cluster can drastically modify its behavior, as illustrated in Figure \Cref{fig:MsdSynthetic} (c). The larger a cluster is, the more likely it is for its center of mass to climb along the c-axis.

\vspace{\MyFigSpace}

\vspace{\MyFigSpace}

\begin{figure}[H]
\centering
\includegraphics[width=0.85\linewidth]{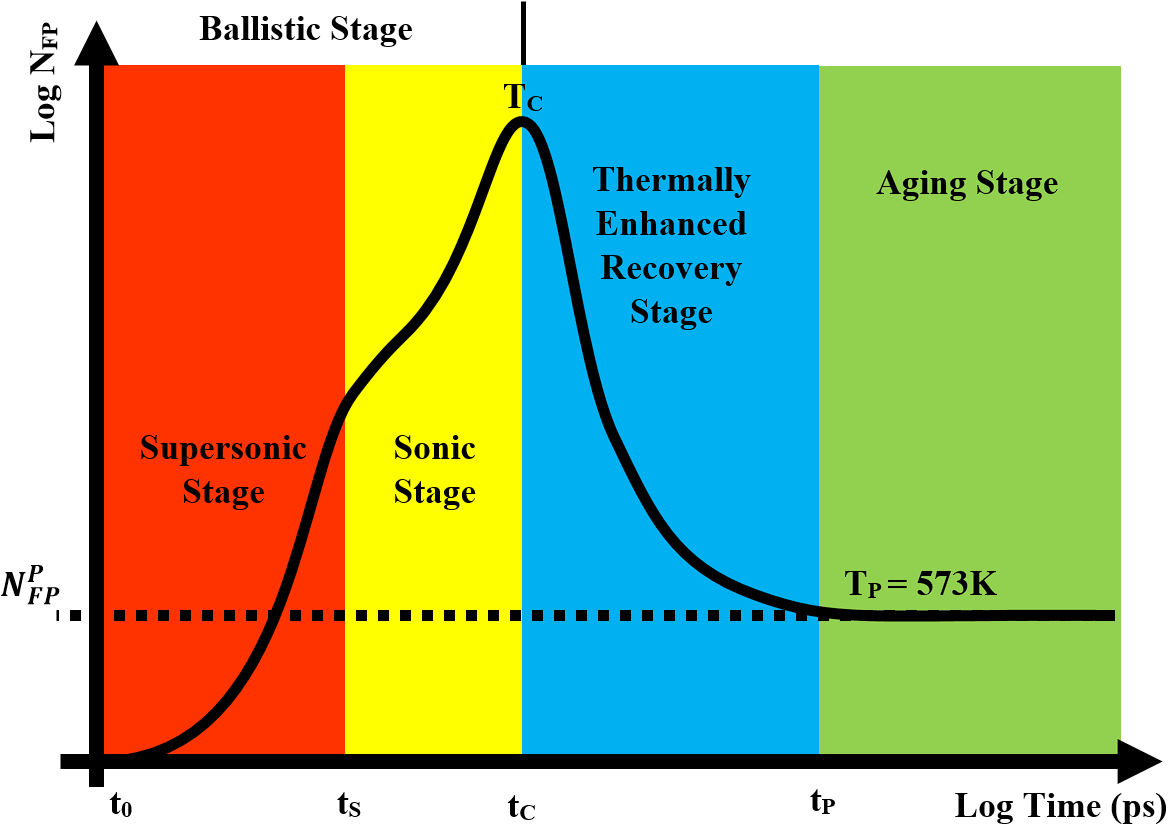}
\vspace{\MyFigSpace}
\caption{Schematic representation of cascade stages in an NFP-Time curve}
% \vspace{\MyFigSpace}
\label{fig:CascadeSchematic}
\end{figure}
\vspace{-2em}

\section{Cascade simulation}
\subsection{Setup}
Using the M2R and M3/M3R potentials, irradiation of pure zirconium was simulated by LAMMPS with PKA energy of 10, 20, 30 and 40 KeV. To achieve statistically significant results, 64 equidistributed PKA directions were tried. We used a box of size (140 × 80 × 60 unit cells) containing 2,688,000 hcp atoms with a lattice constant of 3.23 Å. Unit cells are defined as previously discussed \cref{table:AggregationUnitCell}. Three main stages describe a collision cascade: a supersonic shockwave occurs between times (t$_0$-t$_S$), followed by a sonic shockwave between times (t$_S$-t$_C$), followed by a thermally enhanced recovery stage (t$_C$-t$_P$), as sketched in \cref{fig:CascadeSchematic} \cite{calder2010origin}. In some references, the supersonic and sonic stages are called the ballistic stage  \cite{beland2016effect}. Because of the very high velocities involved during the first two stages, very short timesteps must be used at first during these off-equilibrium MD simulations. As can be seen in \cref{table:CascadeStages}, we first equilibrate (heat) the system to 573K, provide kinetic energy to the PKA, and employ short timesteps during the first few picoseconds of the cascade runs. Furthermore, a variable timestep was employed as a fail-safe (using the dt/reset LAMMPS command), to ensure atomic displacements between each timestep are limited to no more than 0.007 \AA.

% \vspace{-2em}
\begin{table*}[!t]
\caption{LAMMPS setup for simulation of clusters produced by cascade.}
\vspace{\MyFigSpace}
\centering
\begin{tabular}{llllll}
\multicolumn{2}{l}{}                   & Ensemble & Steps       & Dump interval & Max timestep (ps) \\ \hline
1.1. & \multicolumn{1}{l|}{Heating to 573K}   & NPT      & 600,000     & 100,000         & 0.002         \\
1.2. & \multicolumn{1}{l|}{Relaxation to 573K}   & NVE      & 20,000     & 10,000         & 0.002         \\
1.3. & \multicolumn{1}{l|}{Killing Momentum}   & NVE      & 1     & 1         & 0.0001         \\
2. & \multicolumn{1}{l|}{Cascade 1}        & NVE      & 10,000     & 1,000         & 0.0000001         \\
3. & \multicolumn{1}{l|}{Cascade 2}          & NVE      & 50,000           & 5,000             & 0.00001     \\
4. & \multicolumn{1}{l|}{Cascade 3} & Special      & 100,000     & 50           & 0.001
\end{tabular}
% \vspace{\MyFigSpace}
\label{table:CascadeStages}
\vspace{\MyFigSpace}
\end{table*}

% \vspace{-0.5em}

\begin{figure*}[!t]
\centering
\includegraphics[width=0.87\linewidth]{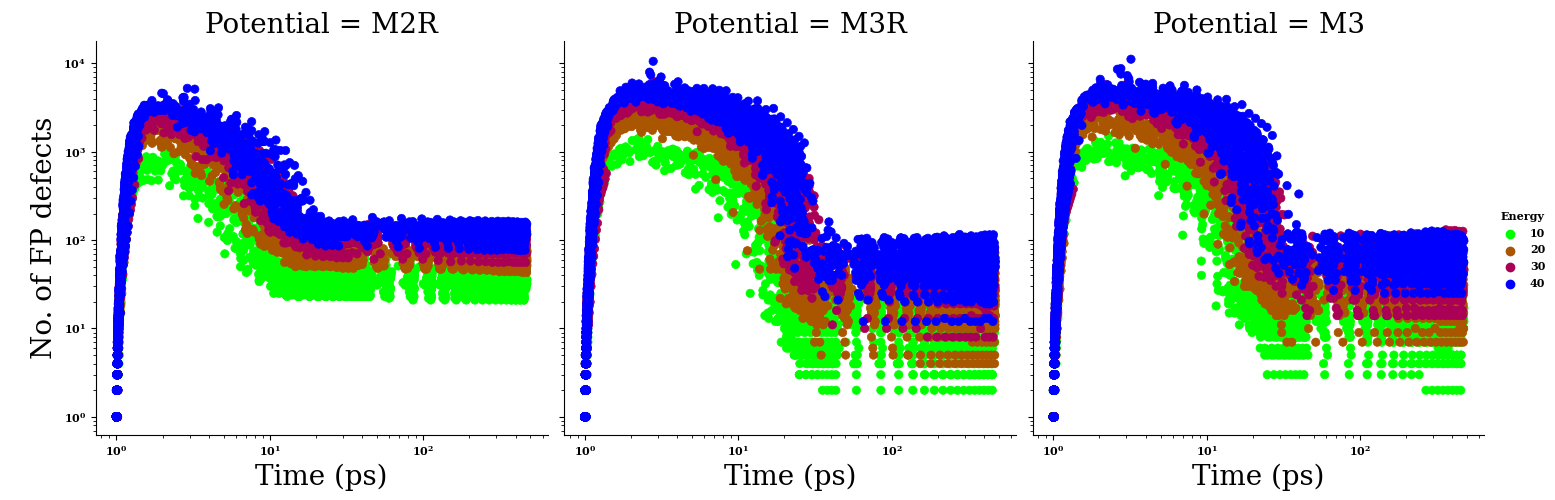}
\vspace{\MyFigSpace}
\caption{Number of FPs for PKA energies of 10,20,30,40 keV.}
\label{fig:FpCount}
\vspace{\MyFigSpace}
\end{figure*}

% \vspace{-0.5em}

\begin{figure*}[!t]
\centering
\includegraphics[width=0.87\linewidth]{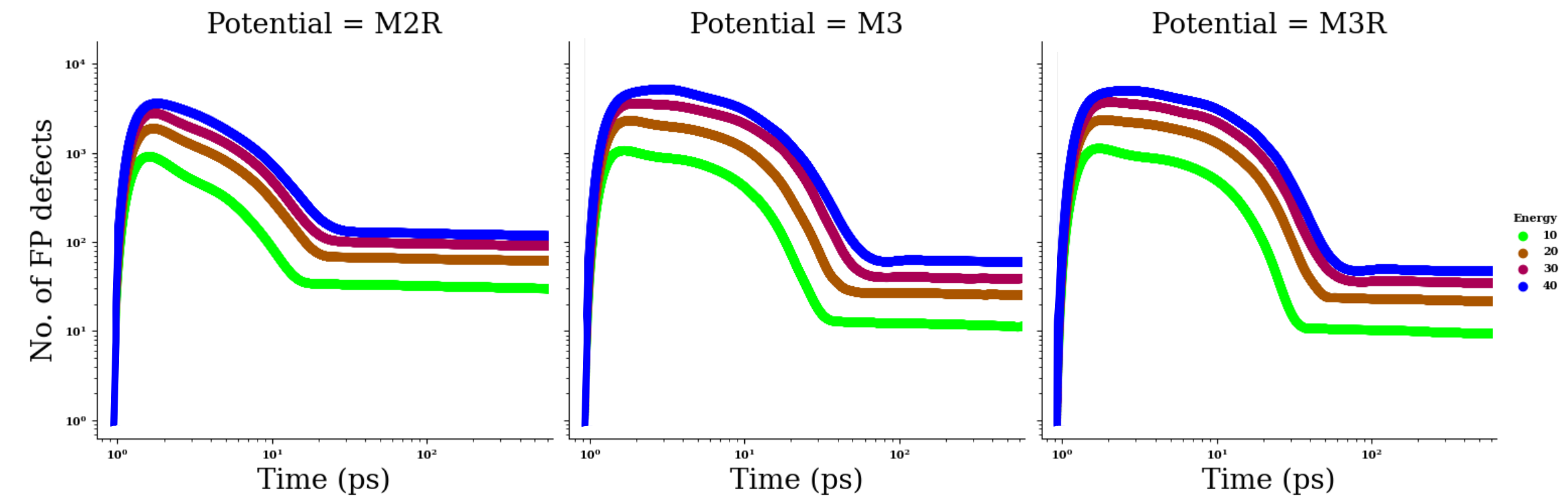}
\vspace{\MyFigSpace}
\caption{Kernel density estimation (KDE) for FP count}
\label{fig:FpKde}
% \vspace{\MyFigSpace}
\end{figure*}

The system was heated from 0K to 573K which is a typical operating temperature for a nuclear power plant \cite{arevalo2007temperature}. Furthermore, this temperature is believed to be a turning point below which vacancy cluster concentration is lower than SIA cluster and beyond it, it is vise versa \cite{griffiths1988review}. The "special" ensemble refers to a mostly un-thermostated simulation, bounded by a 573K Langevin bath of 2 unit cell thickness. A damping factor of 100 was chosen. 

\subsection{Frenkel pairs measurement}
Wigner-Seitz analysis was used in Ovito to determine the number of Frenkel pairs (FP)--i.e. atomic displacements—as illustrated in \cref{fig:FpCount}. This figure confirms that the defect evolution is well converged at the end of the cascade. Furthermore, this figure shows a defect recombination fraction of $\sim$\%78 and $\sim$\%97 for M2R and M3R, respectively. These fractions also slightly depend on PKA energy, and average on \%91. We performed a kernel average of displacements as a function of time using a Gaussian kernel density estimation (KDE) as implemented in SciPy \cite{2020SciPy-NMeth}. These averages are presented in \cref{fig:FpKde}. Comparing M3R to M3 shows that reparameterization had a minor effect on the cascade dynamics and primary damage production. Note that in these models, electron stopping was ignored which can play an important role, especially in our higher-energy simulations \cite{yang2021full}. It is worth mentioning that while M3 is suitable for radiation simulations, M2 is susceptible at short atomic distances, causing errors in some high-energy cascades. 

\subsection{arc-dpa model and M2 cascades}
\begin{figure}[H]
\centering
\includegraphics[width=0.75\linewidth]{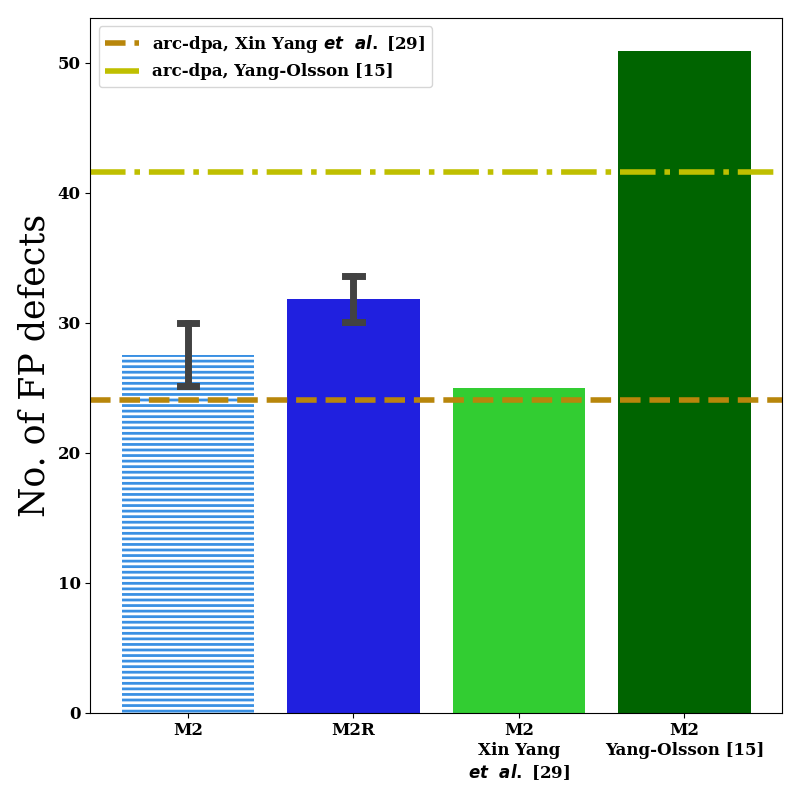}
\vspace{\MyFigSpace}
\caption{The count of FP defects after aging stage compared with arc-dpa estimation.}
\label{fig:ArcDpa}
\vspace{\MyFigSpace}
\end{figure}

\Cref{fig:ArcDpa} shows the average number of remaining FP defects after annealing the cascade debris for 500ps. The average is taken over 64 runs, each involving a 10 keV PKA. This figure also shows two athermal recombination corrected
displacements per atom (arc-dpa) models for comparison, which are both updates to the original model proposed by Nordlund \textit{et al.} \cite{nordlund2018improving}. This model has two material constants, $b_{arc-dpa}=-1$ and $c_{arc-dpa}=0.70\pm0.20$ for Zr, which predicts $\sim$88 FP defects at 10 KeV. The first model uses the parameters suggested by Xin Yang \textit{et al.} \cite{yang2018molecular}. After running cascade simulations based on the M2 potential \cite{mendelev2007development}, Xin Yang \textit{et al.} concluded $b_{arc-dpa}=0.18$ and $c_{arc-dpa}=-0.565$ were a good fit to their MD. Their cascade results and arc-dpa predictions are in agreement with our M2 results, with slight deviations likely attributable to statistical uncertainty.

Another arc-dpa model was proposed by Yang and Olsson \cite{yang2021full} which improves over the original arc-dpa formulation by better treating near-threshold events (low PKA energies). Their arc-dpa model was fitted on cascade simulations using Wilson and Mendelev’s \cite{wilson2015anisotropy} Ni-Zr Finnis–Sinclair (FS) potential, modifying the arc-dpa parameterization proposed in Ref. \cite{konobeyev2017evaluation}. This FS potential was itself a refit of Mendelev \textit{et al.}'s \cite{mendelev2012development} Ni-Zr FS potential (the cross-terms were refit to better capture solid-liquid interfaces). The Zr part of this potential is the M2 potential. Their arc-dpa predicted primary damage production is lower than their M2 MD-based estimate. However, both their arc-dpa predictions and M2 MD-based cascade predictions of primary damage production are considerably larger than our own M2 MD estimates. We are unsure about the origin of this discrepancy. However, given the fairly good agreement between our results and those in Ref. \cite{yang2018molecular}, we believe our M2 cascade simulation setup is sound.  

As can be seen in \Cref{fig:ArcDpa}, M2 and M2R lead to similar estimates of primary damage production. The M2R is slightly softer at short distances, and using it leads to a slightly larger amount of damage produced than using M2. This is a good example of the influence the choice of spline points used to connect an EAM potential to ZBL can have on the outcome of MD cascade simulations. Even slight changes, such as those we applied in this study, can have a measurable impact on primary damage production \cite{beland2016atomistic,beland2017accurate,stoller2016impact}. This difference in the procedure to stiffen the EAM potential explains the discrepancy between the arc-dpa estimate of primary damage production reported in Ref. \cite{yang2018molecular} and our M2R MD cascade simulation results.

% \vspace{\MyFigSpace}

% \vspace{\MyFigSpace}
\begin{figure}[H]
\centering
\includegraphics[width=0.75\linewidth]{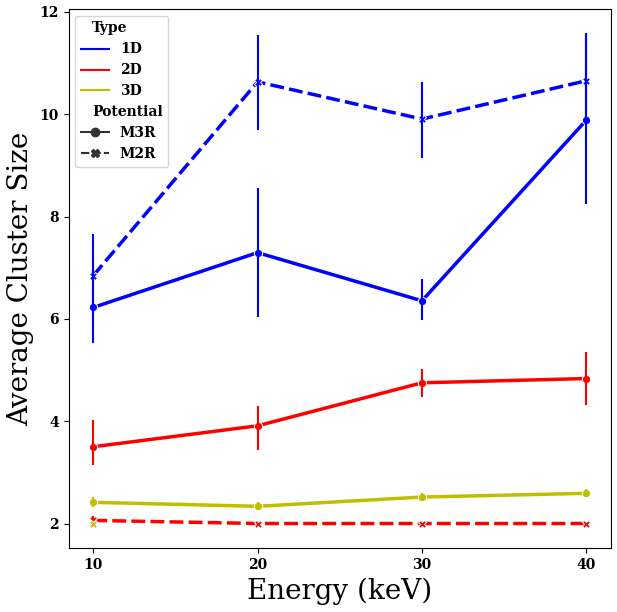}
\vspace{\MyFigSpace}
\caption{Average SIA cluster size}
\label{fig:AverageClusterSize}
\vspace{\MyFigSpace}
\end{figure}

\vspace{\MyFigSpace}
\begin{figure}[H]
\centering
\includegraphics[width=0.9\linewidth]{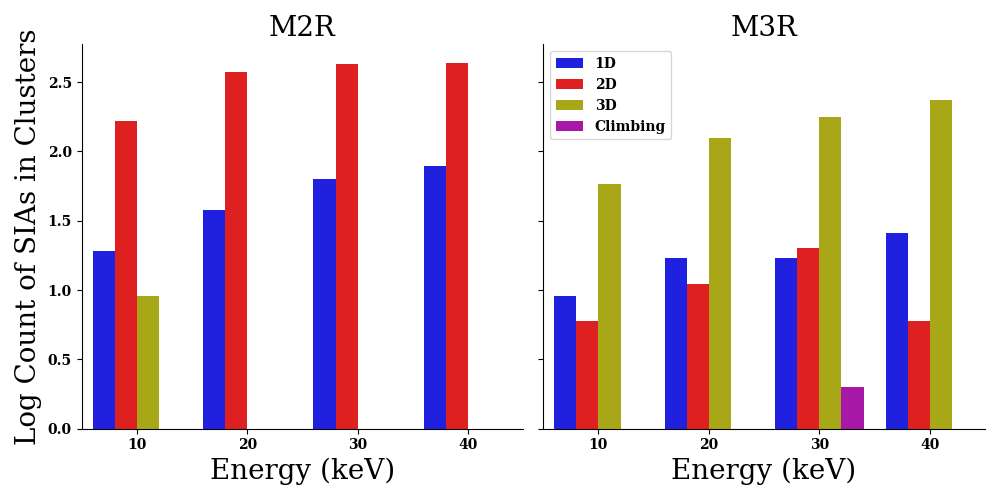}
\vspace{\MyFigSpace}
\caption{No. of SIAs stored as different types of clusters}
\label{fig:CountSiaInCluster}
\vspace{\MyFigSpace}
\end{figure}

\subsection{Cluster Analysis}
Clusters of SIAs were identified throughout the aging simulations. For this analysis, a cutoff distance of 5.8 \AA was set, using the OVITO package.

\subsubsection{Cluster Size}

\paragraph{Visual analysis}
Visual inspection revealed a limited number of pure 1-D diffusing gliders. We observed many 2-D and 3-D clusters. The average cluster size of these clusters has a weak correlation with PKA energy as depicted in \cref{fig:AverageClusterSize}. The total count of SIAs stored as different types of clusters increases with PKA energy, as seen in \cref{fig:CountSiaInCluster}. 

\begin{figure}[!b]
\centering
\includegraphics[width=0.8\linewidth]{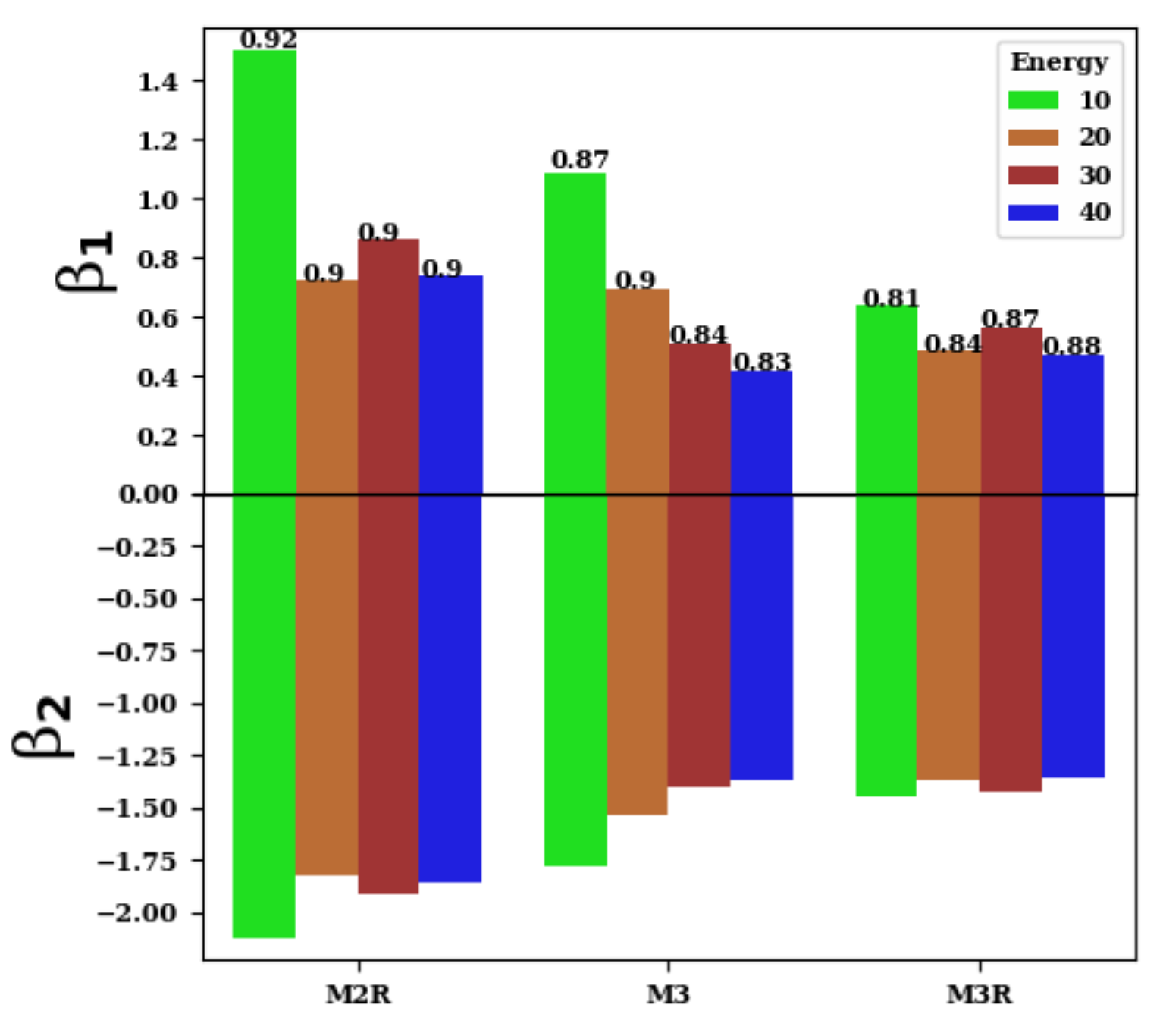}
\vspace{\MyFigSpace}
\caption{Power Law parameters, $\beta_1$ and $\beta_2$, for the model fitted to the cluster size profiles of surviving clusters. The labels above each bar represent the R-squared values.}
\label{fig:PowerLawBeta}
\vspace{\MyFigSpace}
\end{figure}

\begin{figure*}
\centering
\includegraphics[width=\MyFigSize\linewidth]{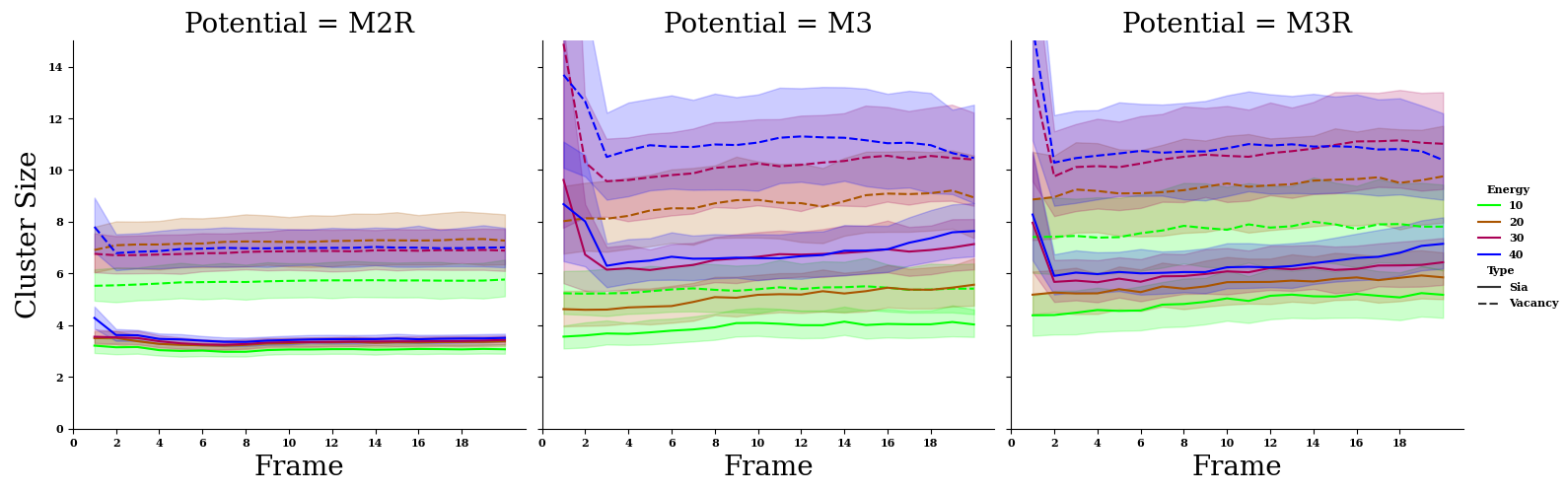}
\vspace{\MyFigSpace}
\caption{Cluster size evolution over a portion of time. The shadows correspond to confidence intervals of 95\%.}
\label{fig:ClusterSize}
\vspace{\MyFigSpace}
\end{figure*}

\begin{figure*}
\centering
\includegraphics[width=\MyFigSize\linewidth]{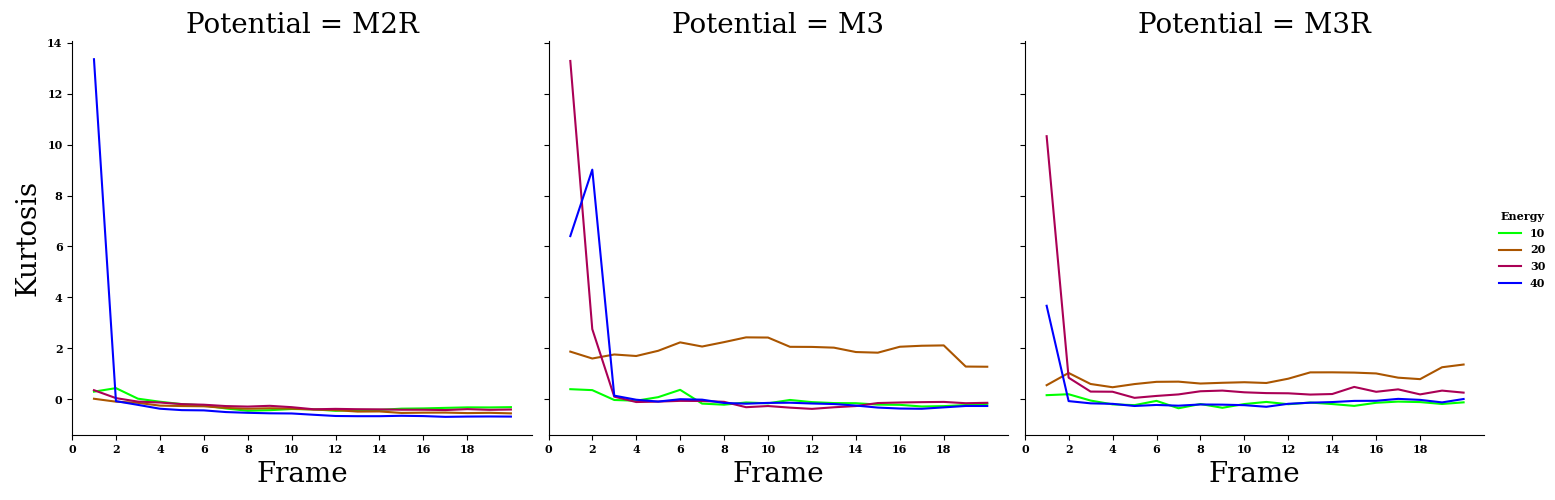}
\vspace{\MyFigSpace}
\caption{Kurtosis of cluster size distribution over time.}
\label{fig:ClusterSizeKurtosis}
\vspace{\MyFigSpace}
\end{figure*}

\begin{figure*}
\centering
\includegraphics[width=\MyFigSize\linewidth]{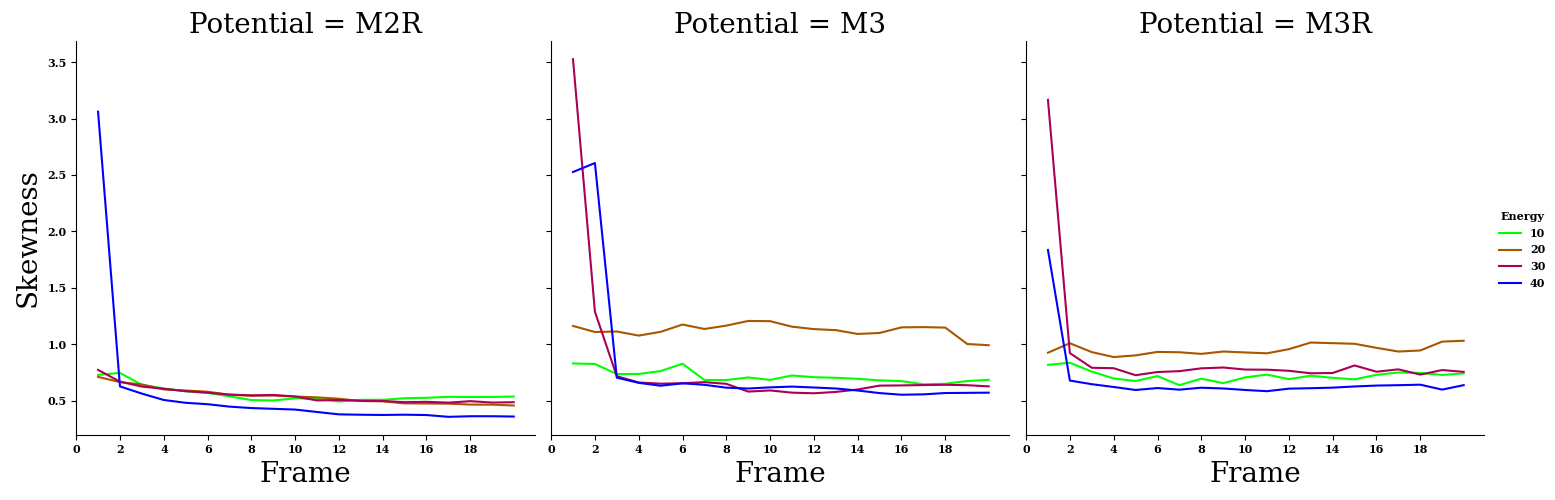}
\vspace{\MyFigSpace}
\caption{Skewness of cluster size distribution over time.}
\label{fig:ClusterSizeSkewness}
\vspace{\MyFigSpace}
\end{figure*}

\begin{figure*}
\centering
\includegraphics[width=\MyFigSize\linewidth]{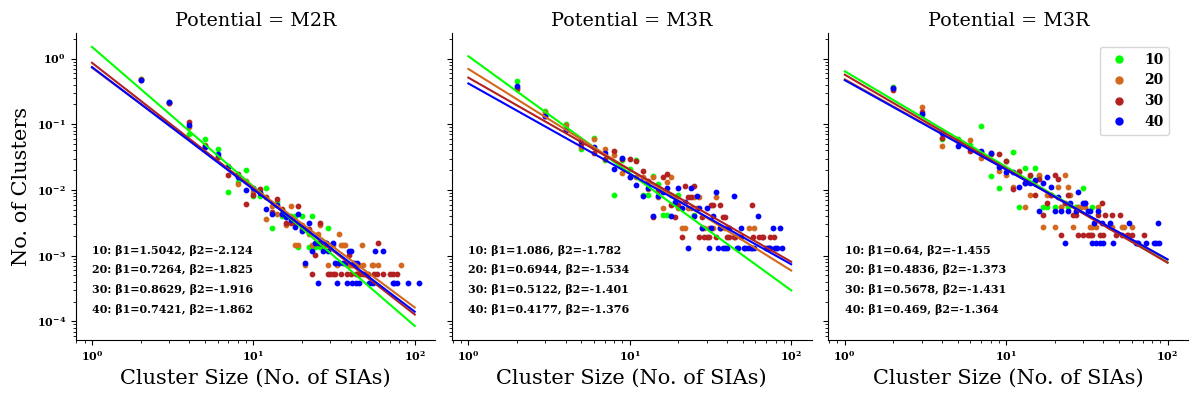}
\vspace{\MyFigSpace}
\caption{Cluster size profile of surviving clusters.}
\label{fig:ClusterSizeProfile}
\vspace{\MyFigSpace}
\end{figure*}

\paragraph{Automated analysis}
A computer program was written to track all of the clusters throughout the aging stage. \Cref{fig:ClusterSize} shows the evolution of cluster size for a portion of this stage. The size distribution of clusters remains almost constant during this period as confirmed by stable kurtosis and skewness depicted in \cref{fig:ClusterSizeKurtosis} and \cref{fig:ClusterSizeSkewness}. \Cref{fig:ClusterSizeProfile} shows the distribution profile of average size for surviving clusters after the aging stage. As expected, smaller clusters are more frequently observed compared to large ones. Also, this figure shows that increasing the PKA energy results in larger SIA clusters. The correlation seen here is close to power law and is probably associated with the fractal nature of collision cascades. Similar power law distributions are reported for different materials \cite{nordlund2019historical,de2018model}. Fitting to a power law equation ($y = \beta_{1}x^{\beta_{2}}$) results in $\beta_1$ and $\beta_2$ as seen in \cref{fig:PowerLawBeta}, where R-squared values are plotted above each bar. 

\subsubsection{Cluster movement analysis}
\paragraph{Visual analysis}
Cluster movement was traced visually, observing the trajectories of defects as identified by Wigner Seitz analysis. We classified the clusters into three groups: 1-D, 2-D and 3D diffusing clusters. 1-D clusters only glide along one of three prismatic directions (either a1 or a2 or a3). 2-D gliders can glide on the basal plane but they do not have any movement along the c direction. These clusters are basically 1-D gliders which can change their gliding direction on the same basal plane from one [11$\bar{2}$0] direction to another. 3D clusters have all components of diffusion (X, Y and Z).

In this work, we did not explore the full energy landscape accessible to these clusters, which likely possess polymorphism (similarly to the case of SIA clusters in bcc Fe \cite{marinica2012irradiation}), with some of the configurations only accessible if long waiting times - i.e. large activation barriers - are considered. The use of techniques such as the activation relaxation technique nouveau (ARTn) \cite{mousseau2012activation} could help better understand these structures and their kinetics. That being stated, this observation is consistent with one of our main conclusions: the pathway by which clusters have been generated matters. 

Cascade-generated clusters have a higher propensity to be structured as crowdions — similarly to a-loops — than those generated by agglomeration. This is consistent with the fact they exhibit more anisotropic diffusion along the a-axis. However, we should emphasize that these were not dislocation loops: the clusters considered in our work were too small to truly be considered as such; dislocation line segments could not be clearly identified.

% \clearpage
\vspace{-0.1em}

\begin{figure*}[!t]
    \centering
        \begin{minipage}{.3\textwidth}
            \centering
            \begin{subfigure}{\textwidth}
                \centering
                \vspace{\MyFigSpace}
                \caption*{Potential = M2R}
                \includegraphics[width=\MyFigSize\textwidth]{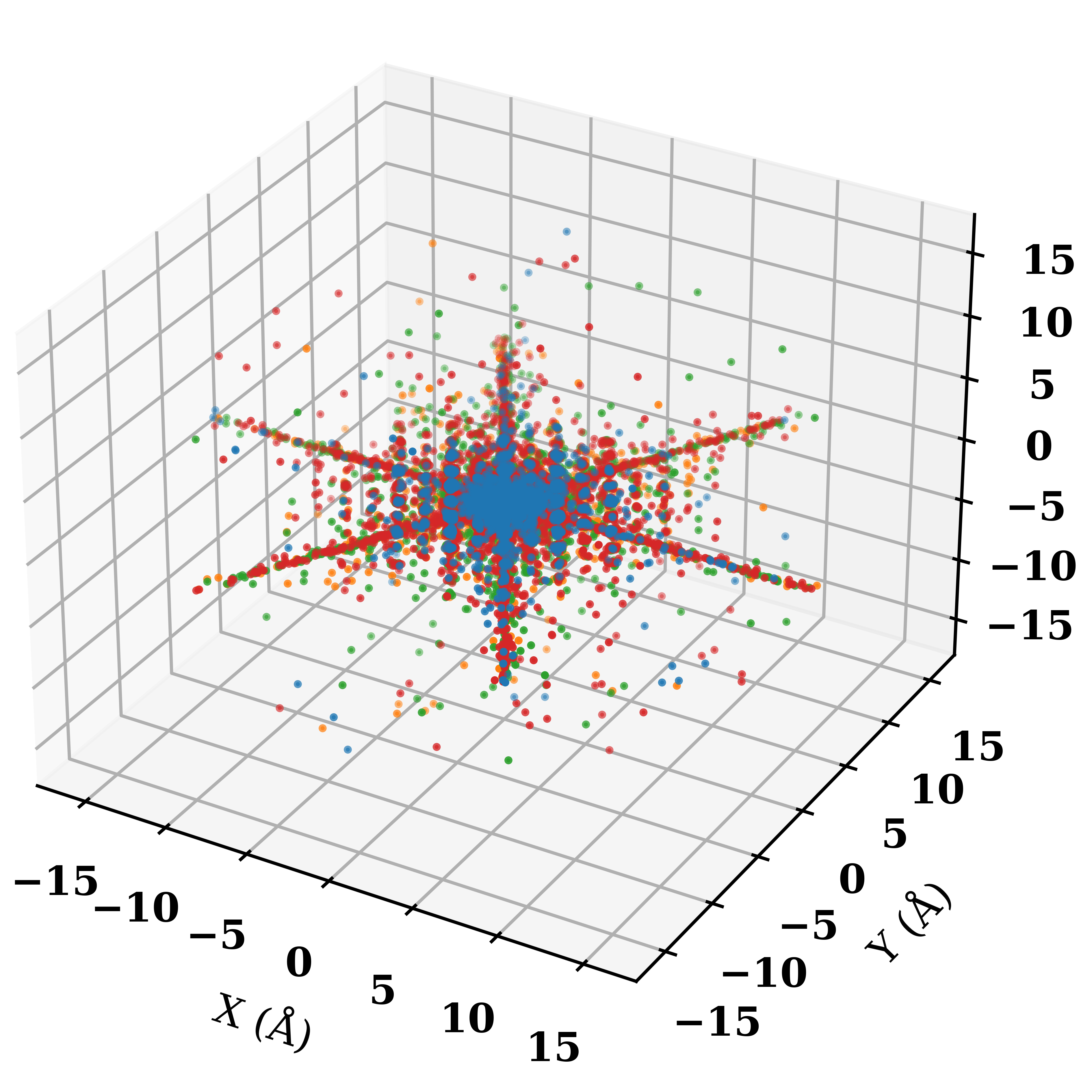}
                \vspace{\MyFigSpace}
            \end{subfigure}%
            % \caption*{Column 1}
            \label{fig:col1}
        \end{minipage}
        % \hfill
        \begin{minipage}{.3\textwidth}
            \centering
            \begin{subfigure}{\textwidth}
                \centering
                \vspace{\MyFigSpace}
                \caption*{Potential = M3}
                \includegraphics[width=\MyFigSize\textwidth]{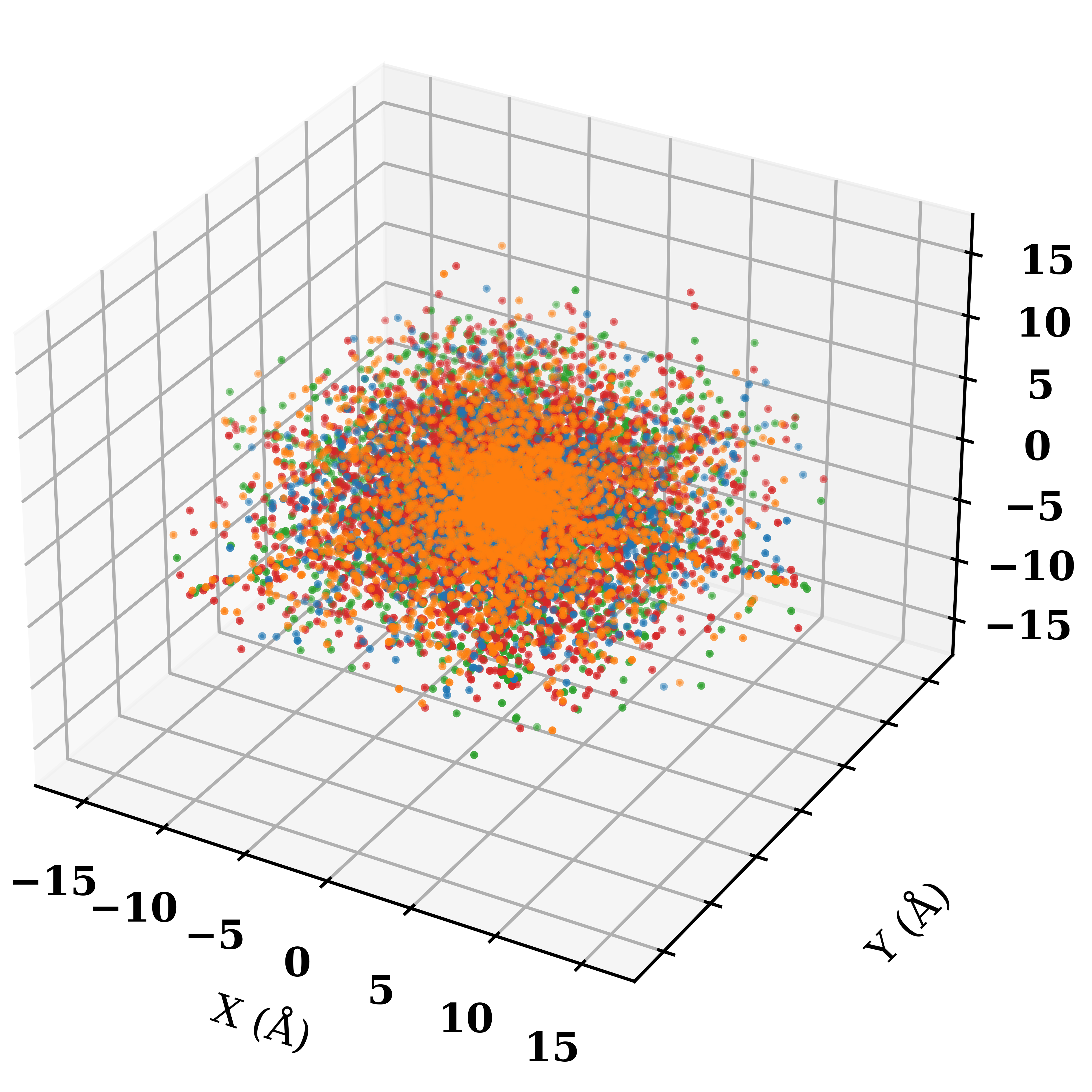}
                \vspace{\MyFigSpace}
            \end{subfigure}%\\[\baselineskip]
        \end{minipage}
        % \hfill
        \begin{minipage}{.3\textwidth}
            \centering
            \begin{subfigure}{\textwidth}
                \centering
                \vspace{\MyFigSpace}
                \caption*{Potential = M3R}
                \includegraphics[width=\MyFigSize\textwidth]{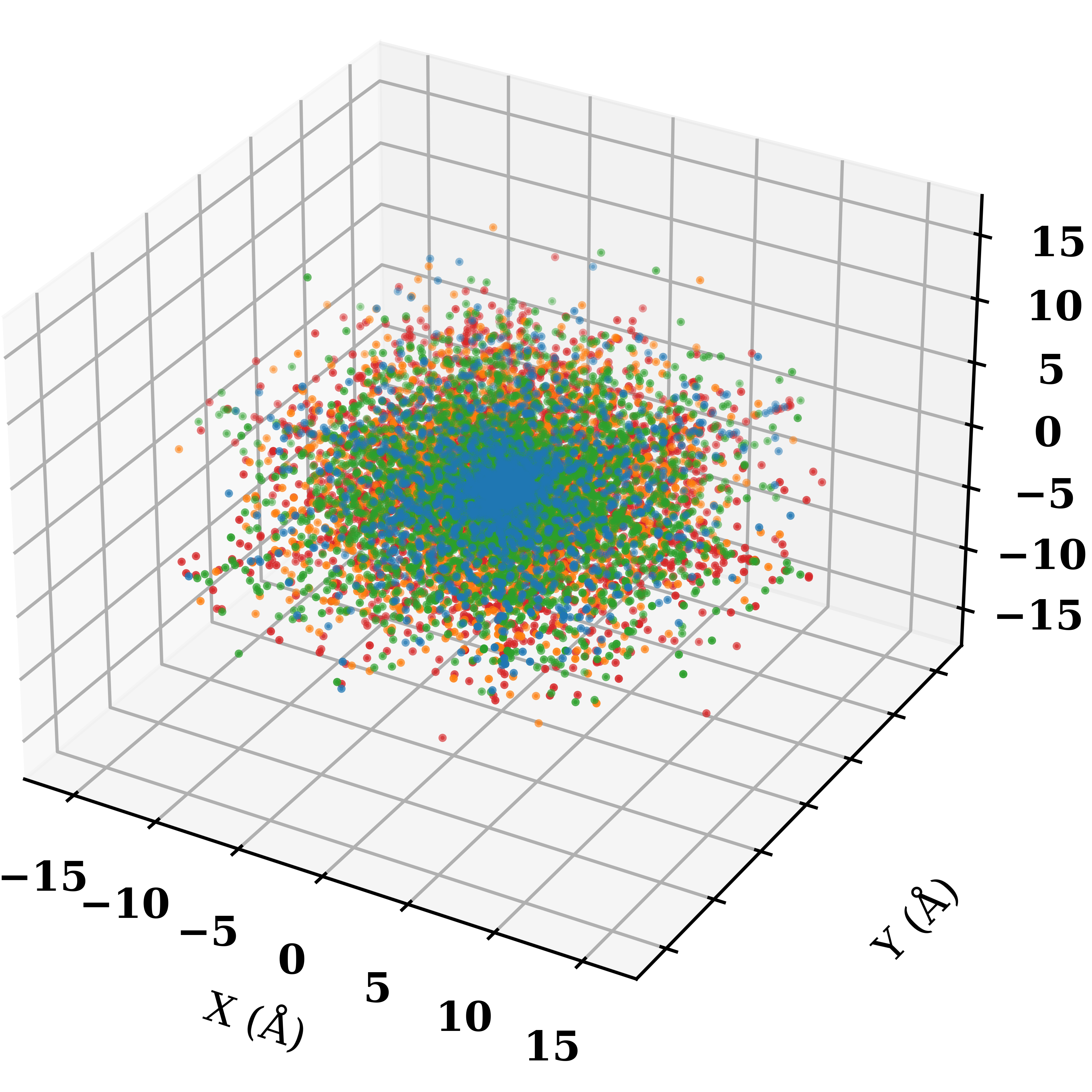}
                \vspace{\MyFigSpace}
            \end{subfigure}%\\[\baselineskip]
        \end{minipage}
    \caption{3D displacement vectors for cluster movement in the aging stage.}
    \label{fig:ClusterMovement3D}
\end{figure*}

\begin{figure*}[!t]
    \centering
        \begin{minipage}{.3\textwidth}
            \centering
            \begin{subfigure}{\textwidth}
                \centering
                \vspace{-2em}
                \caption*{Potential = M2R}
                \includegraphics[width=\MyFigSize\textwidth]{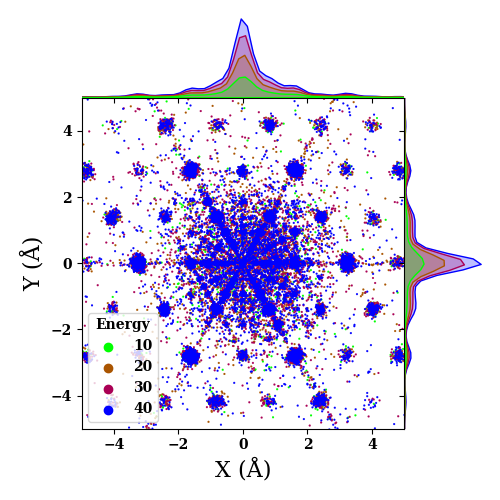}
                \vspace{\MyFigSpace}
            \end{subfigure}%
            % \caption*{Column 1}
            \label{fig:col1}
        \end{minipage}
        % \hfill
        \begin{minipage}{.3\textwidth}
            \centering
            \begin{subfigure}{\textwidth}
                \centering
                \vspace{-2em}
                \caption*{Potential = M3}
                \includegraphics[width=\MyFigSize\textwidth]{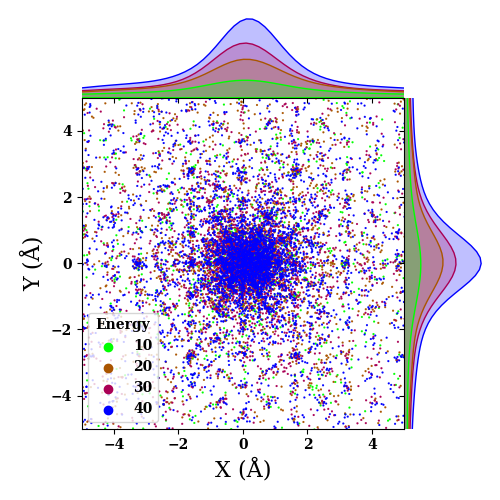}
                \vspace{\MyFigSpace}
            \end{subfigure}%\\[\baselineskip]
        \end{minipage}
        % \hfill
        \begin{minipage}{.3\textwidth}
            \centering
            \begin{subfigure}{\textwidth}
                \centering
                \vspace{-2em}
                \caption*{Potential = M3R}
                \includegraphics[width=\MyFigSize\textwidth]{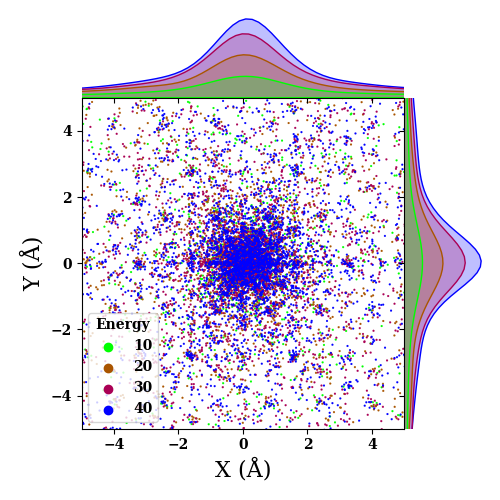}
                \vspace{\MyFigSpace}
            \end{subfigure}%\\[\baselineskip]
        \end{minipage}
    \caption{projection of displacement vectors on basal plane.}
    \label{fig:ClusterMovementProjection}
\end{figure*}

\begin{figure*}[!t]
\centering
\includegraphics[width=\MyFigSize\linewidth]{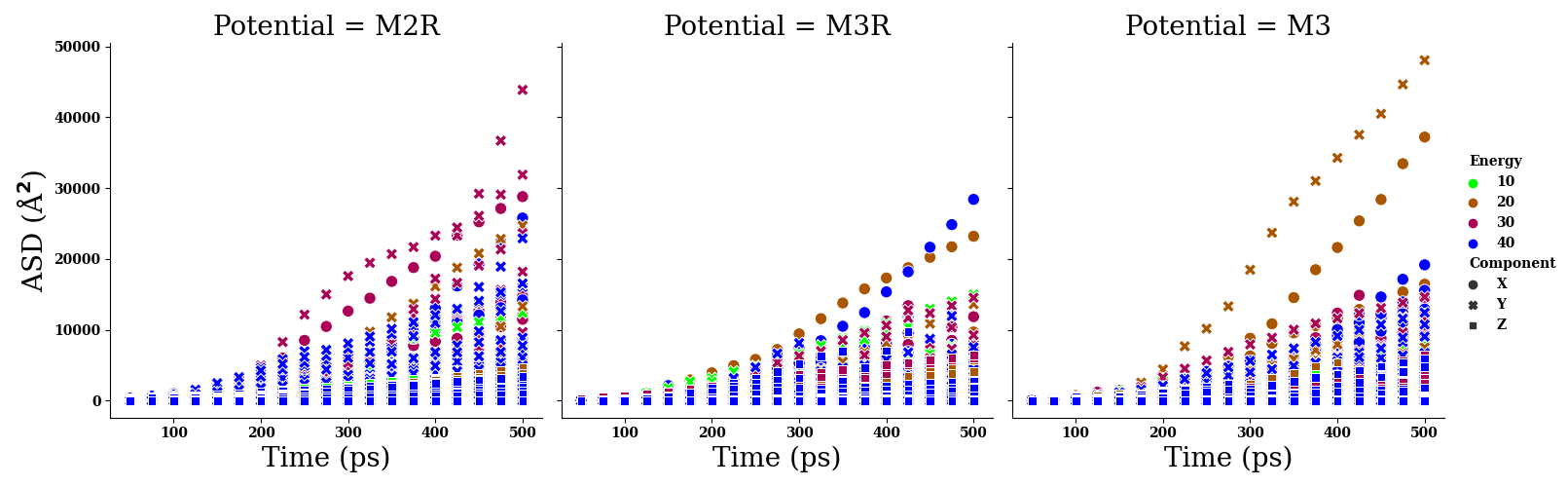}
\vspace{\MyFigSpace}
\caption{Square displacement (SD) of clusters during post-cascade aging}
\label{fig:Sd}
\vspace{\MyFigSpace}
\end{figure*}

\begin{figure*}[!t]
\centering
\includegraphics[width=\MyFigSize\linewidth]{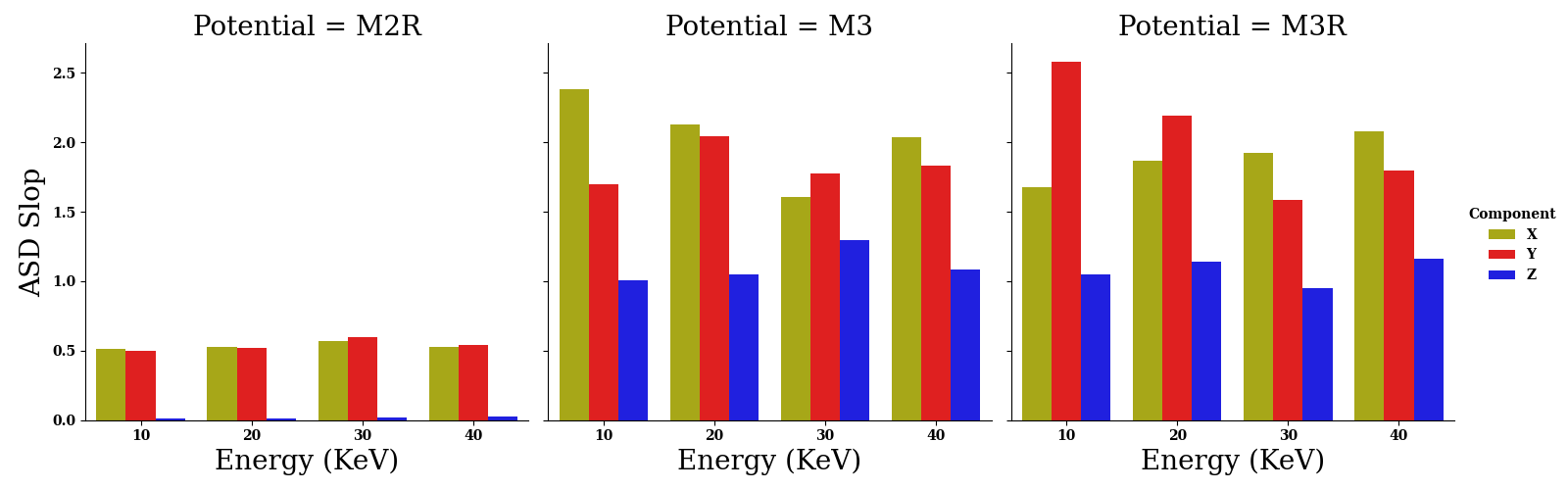}
\vspace{\MyFigSpace}
\caption{Slope of SD displacements of the clusters}
\label{fig:AsdSlope}
\vspace{\MyFigSpace}
\end{figure*}

\begin{figure*}[!t]
\centering
\includegraphics[width=0.9\linewidth]{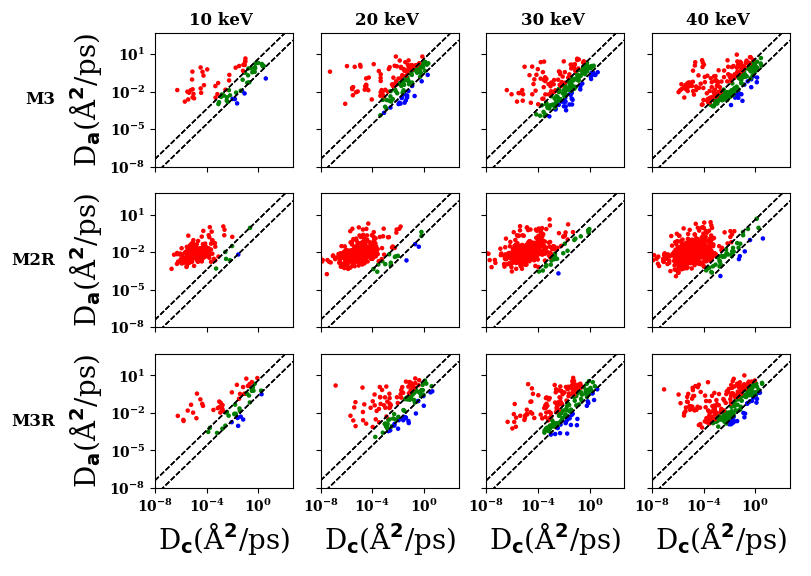}
\caption{Dc vs. Da scatter profile of cascade-induced clusters.}
\label{fig:DcDaScatter}
\end{figure*}

\begin{figure*}[!t]
\centering
\includegraphics[width=\MyFigSize\linewidth]{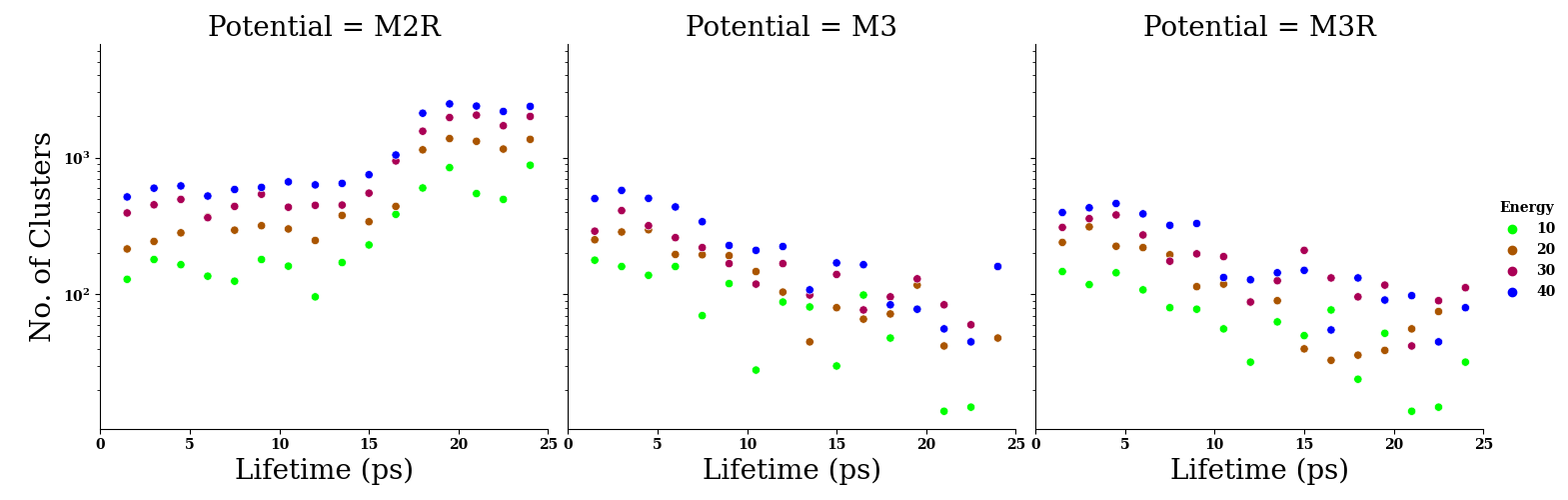}
\caption{Lifetime profile of glissile clusters.}
\label{fig:LifeTime}
\end{figure*}

\begin{figure} [H]
\centering
\includegraphics[width=0.7\linewidth]{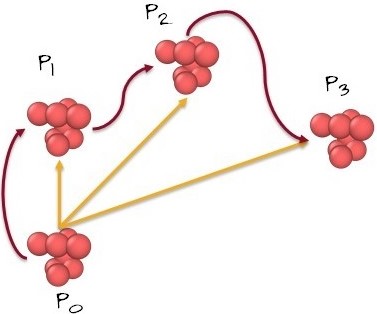}
% \vspace{\MyFigSpace}
\caption{Schematic depiction of how cluster movement vector is measured compared to its initial position.}
\label{fig:ClusterMovementSchematic}
\end{figure}

\begin{figure}[H]
\centering
\includegraphics[width=0.99\linewidth]{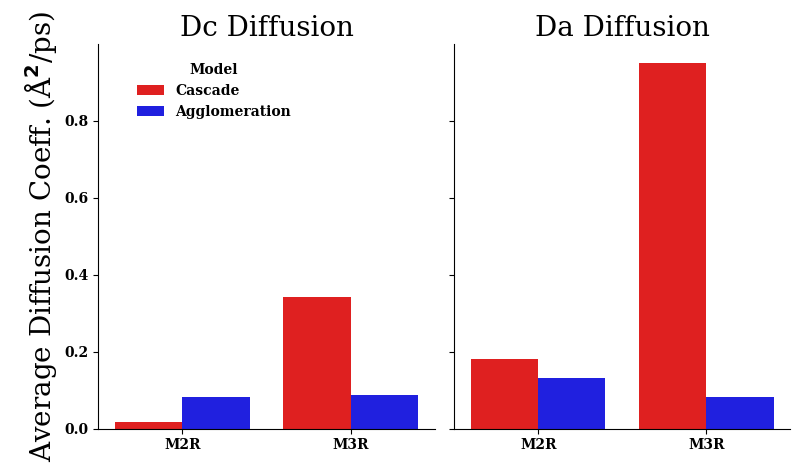}
\vspace{\MyFigSpace}
\caption{Average diffusion coefficients of clusters created by agglomeration of single SIAs in comparison with cascade-induced clusters.}
\vspace{\MyFigSpace}
\label{fig:DiffCoeff}
\end{figure}

\paragraph{Automated analysis}
Our program first identified clusters based on the cluster size and position of the center of gyration. Next, the displacement vector of each cluster was measured compared to the previous frame as depicted in \cref{fig:ClusterMovementSchematic}. \Cref{fig:ClusterMovement3D} illustrates these movement vectors in 3D space. \Cref{fig:ClusterMovementProjection} shows the projection of displacement vector on the basal plane. As is seen in this graph, cluster glide occurs mostly along a1, a2 and a3. After identification and measuring displacement vectors, square displacement (SD) was calculated. \Cref{fig:Sd} shows SD in the aging stage for some of the runs where a few clusters stand out due to their large gliding bias. Next, the slopes of SD curves were averaged, as shown in \cref{fig:AsdSlope}, and used to calculate the diffusion coefficients along basal and prismatic direction based on \cref{SdEquation,EqDc,EqDa}. \Cref{fig:DcDaScatter} shows Dc vs. Da for these clusters. \Cref{fig:DiffCoeff} shows the average of within-basal-plane (Da) and out-of-plane (Dc) diffusion for cascade-induced clusters in comparison with agglomerated clustering. This higher diffusion rate is an indication severely out-of-equilibrium conditions can lead to different SIA cluster structures as compared to near-equilibrium processes. The surrounding debris and their stress fields may also affect cluster diffusion coefficients. We report the lifetime profile of these glissile clusters in \cref{fig:LifeTime}. As is seen, glissile clusters tend to linger on for longer periods of time without interacting with other debris when M2R is used, compared to M3 and M3R. This is mainly due to the fact that M3 and M3R create faster diffusing clusters, as seen in \cref{fig:DiffCoeff}, in spite of their less damage production. Such interactions will alter the nature of the cluster, resetting the lifetime clock.

\begin{equation}
SD_{x,y,z}=R_{cl}\sum_{}^{}(p_{(t)}-p_{(0)})^{2}
\label{SdEquation}
\end{equation}

\begin{equation}
D_{c}=\frac{SD_{z}}{4t_{int}}
\label{EqDc}
\end{equation}

\begin{equation}
D_{a}=\frac{SD_{x}+SD_{y}}{4t_{int}}
\label{EqDa}
\end{equation}

\begin{figure}
    \begin{tikzpicture}
    \node(a){\includegraphics[width=0.9\linewidth]{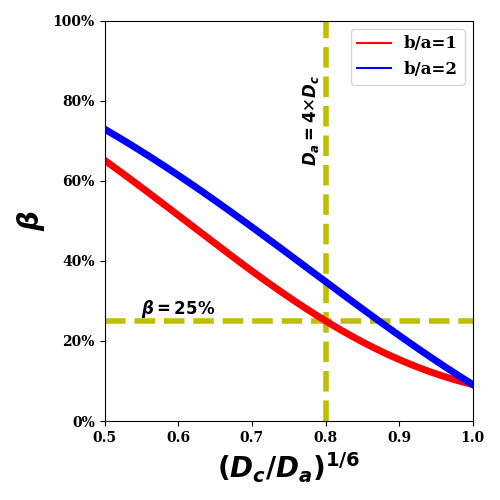}};
    \node at (a.north east)
    [
    anchor=center,
    xshift=-16mm,
    yshift=-32mm
    ]
    {
        \includegraphics[width=0.12\textwidth]{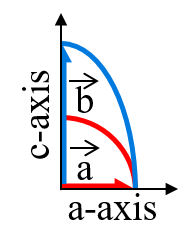}
    };
    \end{tikzpicture}
\vspace{\MyFigSpace}
\caption{Defining the criteria for categorizing gliders using the notion of bias factor.}
\vspace{\MyFigSpace}
\label{fig:BiasFactor}
\end{figure}

\section{Calculation of bias factor}

In our simulations, we rarely observed 1-D clusters that glide solely along a single compact basal [11$\bar{2}$0] direction. Instead, these fast diffusing clusters often reconfigure and diffuse along another compact basal direction on the same plane, forming 2-D clusters. Over time, they occasionally climb along the c-axis and then glide on another basal plane. This preferential glide creates a biased flow of clusters, which we refer to as a-biased diffusing clusters (ADC). To differentiate ADCs from isotropically diffusing clusters (IDC) and c-biased diffusing clusters (CDC), which preferentially diffuse along the prismatic direction.

We propose using the bias factor ($\beta$) formulation that was initially developed by Woo in 1988 \cite{woo1988theory} and later was modified by Saidi \textit{et al.} \cite{saidi2020method} as this criterion. The bias factor in this formulation has two parts: a constant bias due to elastic interaction (10\%) and a part that depends on the orientation, geometry and extension of the sinks with respect to prismatic and basal planes. The bias factor due to sink orientation, depends on the anisotropy of diffusion. The anisotropy of diffusion is represented by the p-factor that is the 6$^{th}$ root of D$_c$/D$_a$. According to this formulation, a bias factor of 25\% for a fully circular dislocation loop (b/a=1) necessitates a p-factor of 0.8, as shown in \cref{fig:BiasFactor}, which is equivalent to D$_a$ being four times larger than D$_c$. Naturally, a more conservative criterion (higher $\beta$) necessitates a lower p-factor (lower D$_c$/D$_a$ ratio) which in turn expands the isotropic region (green) in \cref{fig:DcDaScatter} and lowers $\varepsilon_{i}^{g}$.

\section{Epsilon}
\subsection{Visual analysis}
\begin{figure*}
    \centering
        \begin{minipage}{.38\textwidth}
            \centering
            \begin{subfigure}{\textwidth}
                \centering
                \caption*{Size \geq 2 SIAs}
                \caption*{(a)}
                \makebox[0pt][r]{\makebox[40pt]{\raisebox{95pt}{\rotatebox[origin=c]{90}{$1-D$}}}}%
                \includegraphics[width=\textwidth]{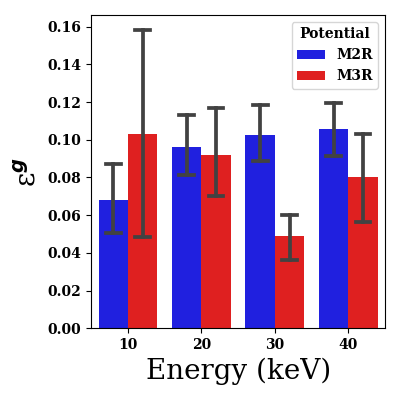}
                \vspace{-2em}
            \end{subfigure}%\\[\baselineskip]
                \begin{subfigure}{\textwidth}
                \centering
                \caption*{(c)}
                \makebox[0pt][r]{\makebox[40pt]{\raisebox{95pt}{\rotatebox[origin=c]{90}{$2-D$}}}}%
                \includegraphics[width=\textwidth]{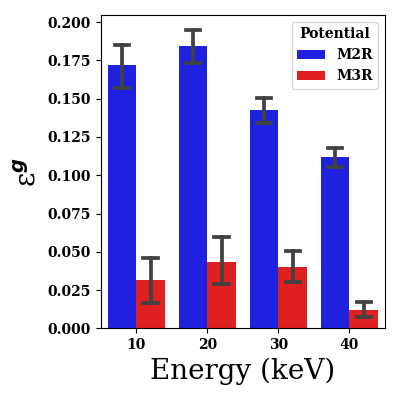}
                \vspace{-2em}
            \end{subfigure}%\\[\baselineskip]
            \begin{subfigure}{\textwidth}
                \centering
                \caption*{(e)}
                \makebox[0pt][r]{\makebox[40pt]{\raisebox{95pt}{\rotatebox[origin=c]{90}{$ADC$ $ $ $(1-D $ $ \& $ $ 2-D)$}}}}%
                \includegraphics[width=\textwidth]{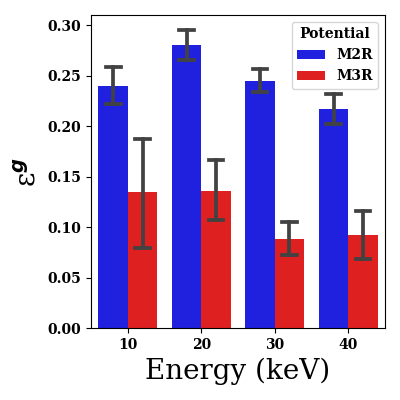}
                \vspace{-2em}
            \end{subfigure}%
            % \caption*{Column 1}
            \label{fig:col1}
        \end{minipage}
        % \hfill
        \begin{minipage}{.38\textwidth}
            \centering
            \begin{subfigure}{\textwidth}
                \centering
                \caption*{Size \geq 3 SIAs}
                \caption*{(b)}
                \includegraphics[width=\textwidth]{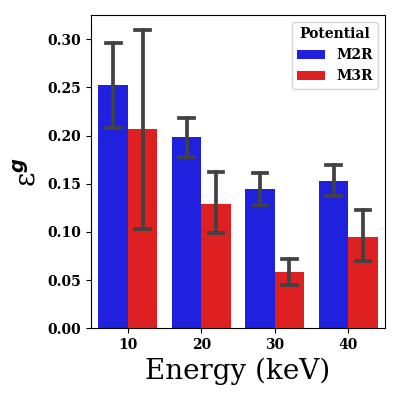}
                \vspace{-2em}
            \end{subfigure}%\\[\baselineskip]
            \begin{subfigure}{\textwidth}
                \centering
                \caption*{(d)}
                \includegraphics[width=\textwidth]{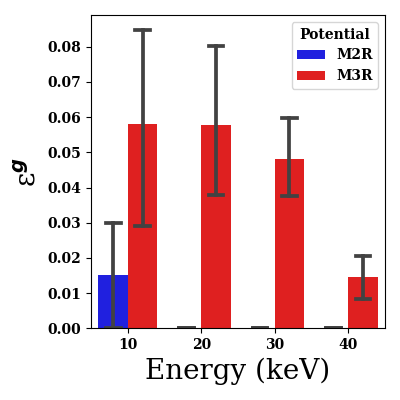}
                \vspace{-2em}
            \end{subfigure}%\\[\baselineskip]
            \begin{subfigure}{\textwidth}
                \centering
                \caption*{(f)}
                \includegraphics[width=\textwidth]{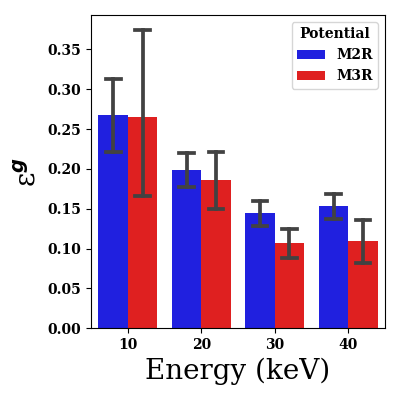}
                \vspace{-2em}
            \end{subfigure}%
            % \caption*{Column 2}
            \label{fig:col2}
        \end{minipage}
    \caption{The effect of minimum cluster size and cluster type on epsilon (based on visual analysis).}
    \label{fig:EpsilonVisual}
\end{figure*}

For 1-D, 2-D and ADCs (1-D plus 2-D) the $\varepsilon_{i}^{g}$ value is estimated based on the clusters remaining at the end of our annealing stage. These estimates are illustrated in \cref{fig:EpsilonVisual}. This figure also shows the $\varepsilon_{i}^{g}$ if one excludes di-SIAs from the calculation (as different force-fields provide widely different estimates of di-SIA anisotropy of diffusion and propensity for cascades to generate 2-SIAs).  If one considers both 1-D and 2-D diffusers contribute to epsilon, the values we obtain for clusters of 3 SIAs and above are broadly in line with the epsilon parameter used by Barashev \textit{et al.} \cite{barashev2015theoretical}. (0.2). If di-SIAs are considered, M2R is associated with a larger value (.24) and M3R is associated with a lesser value (0.13).

\begin{figure*}
    \centering
        \begin{minipage}{.48\textwidth}
            \centering
            \begin{subfigure}{\textwidth}
                \centering
                \caption*{\geq 2 SIAs}
                \caption*{(a)}
                \makebox[0pt][r]{\makebox[30pt]{\raisebox{130pt}{\rotatebox[origin=c]{90}{$D_{a}/C_{c}=4$}}}}%
                \includegraphics[width=\textwidth]{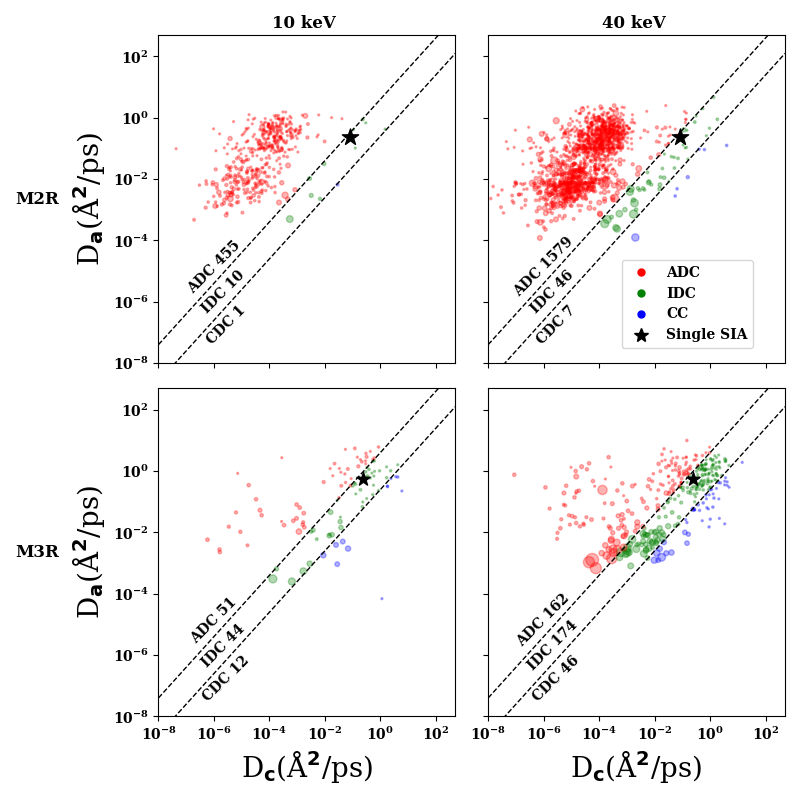}
                \vspace{-2em}
            \end{subfigure}%\\[\baselineskip]
                \begin{subfigure}{\textwidth}
                \centering
                \caption*{(c)}
                \makebox[0pt][r]{\makebox[30pt]{\raisebox{130pt}{\rotatebox[origin=c]{90}{$D_{a}/C_{c}=10$}}}}%
                \includegraphics[width=\textwidth]{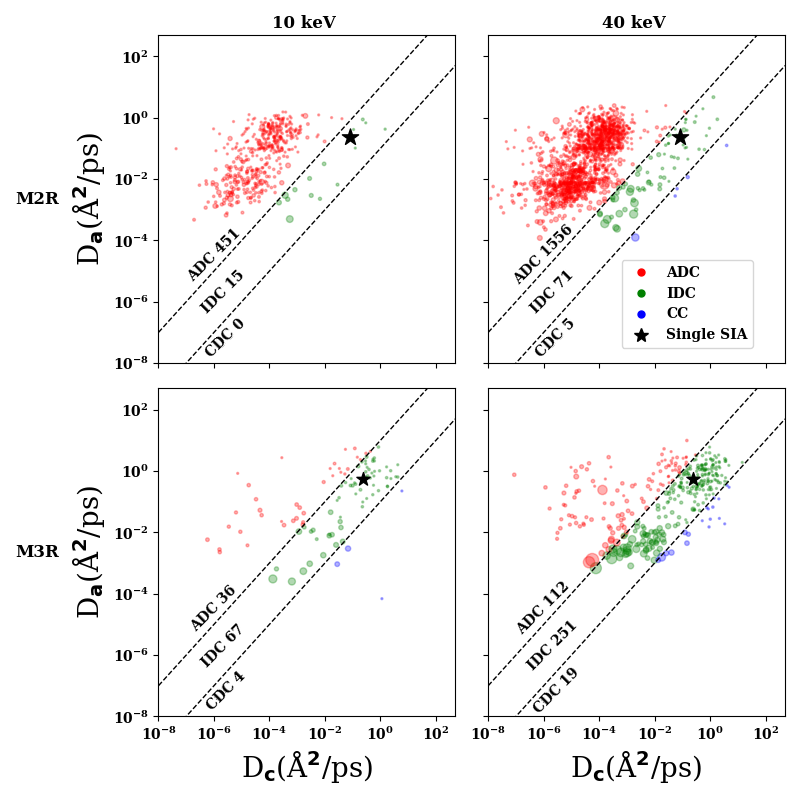}
                \vspace{-2em}
            \end{subfigure}%\\[\baselineskip]
            % \caption*{Column 1}
            \label{fig:col1}
        \end{minipage}
        % \hfill
        \begin{minipage}{.48\textwidth}
            \centering
            \begin{subfigure}{\textwidth}
                \centering
                \caption*{\geq 3 SIAs}
                \caption*{(b)}
                \includegraphics[width=\textwidth]{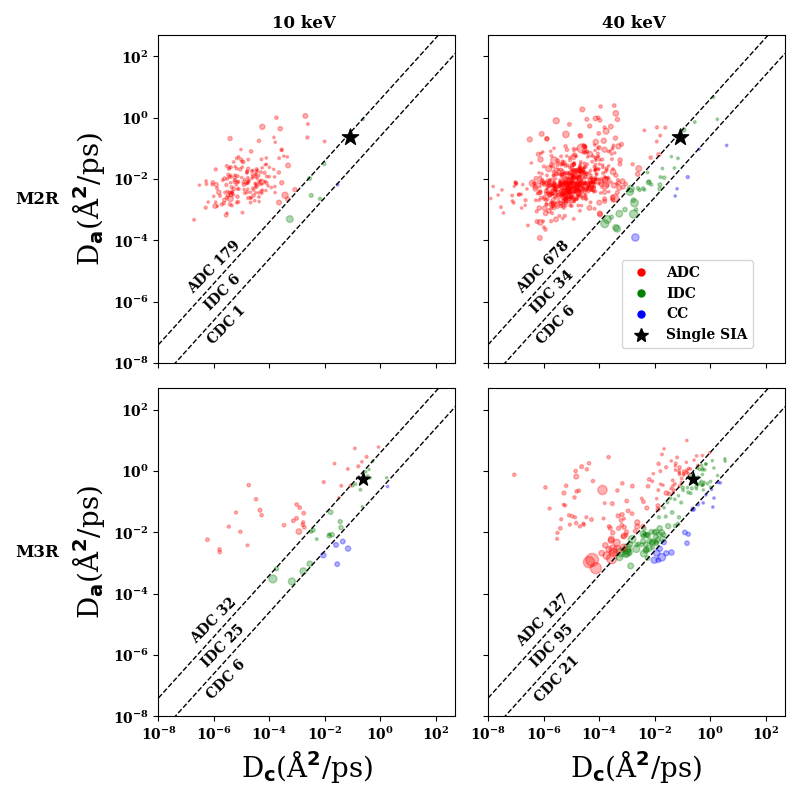}
                \vspace{-2em}
            \end{subfigure}%\\[\baselineskip]
            \begin{subfigure}{\textwidth}
                \centering
                \caption*{(d)}
                \includegraphics[width=\textwidth]{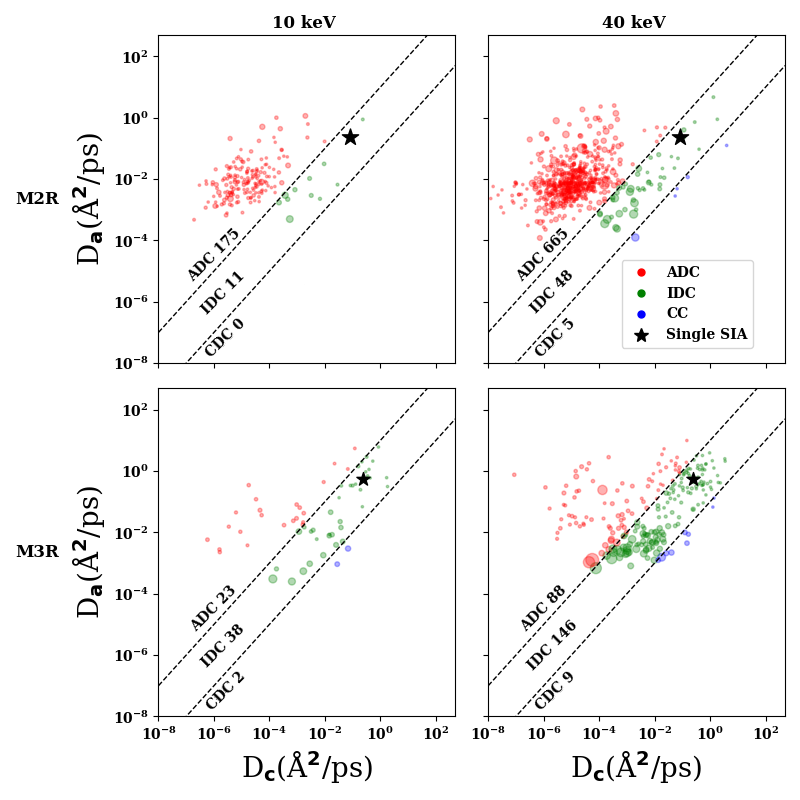}
                \vspace{-2em}
            \end{subfigure}%\\[\baselineskip]
            % \caption*{Column 2}
            \label{fig:col2}
        \end{minipage}
    \caption{The effect of minimum cluster size considered and chosen bias factor on availability and categorization of clusters in automatic analysis}
    \label{fig:DcDaAutomated}
\end{figure*}

\begin{figure*}
    \centering
        \begin{minipage}{.48\textwidth}
            \centering
            \begin{subfigure}{\textwidth}
                \centering
                \caption*{\geq 2 SIAs}
                \caption*{(a)}
                \makebox[0pt][r]{\makebox[30pt]{\raisebox{130pt}{\rotatebox[origin=c]{90}{$D_{a}/C_{c}=4$}}}}%
                \includegraphics[width=\textwidth]{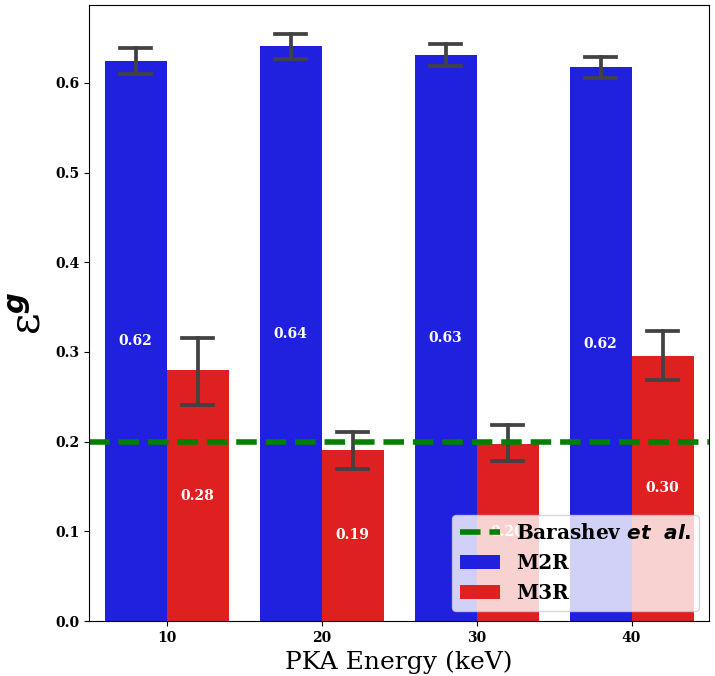}
                \vspace{-2em}
            \end{subfigure}%\\[\baselineskip]
                \begin{subfigure}{\textwidth}
                \centering
                \caption*{(c)}
                \makebox[0pt][r]{\makebox[30pt]{\raisebox{130pt}{\rotatebox[origin=c]{90}{$D_{a}/C_{c}=10$}}}}%
                \includegraphics[width=\textwidth]{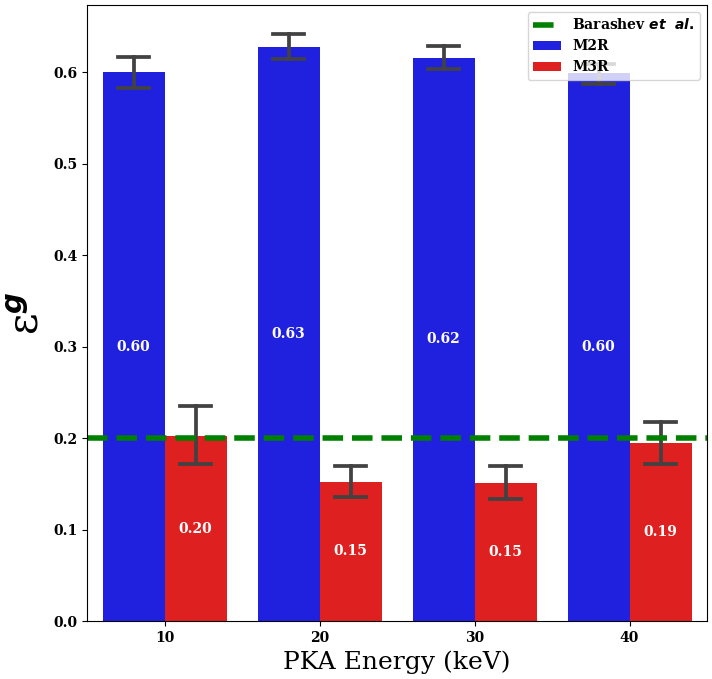}
                \vspace{-2em}
            \end{subfigure}%\\[\baselineskip]
            % \caption*{Column 1}
            \label{fig:col1}
        \end{minipage}
        % \hfill
        \begin{minipage}{.48\textwidth}
            \centering
            \begin{subfigure}{\textwidth}
                \centering
                \caption*{\geq 3 SIAs}
                \caption*{(b)}
                \includegraphics[width=\textwidth]{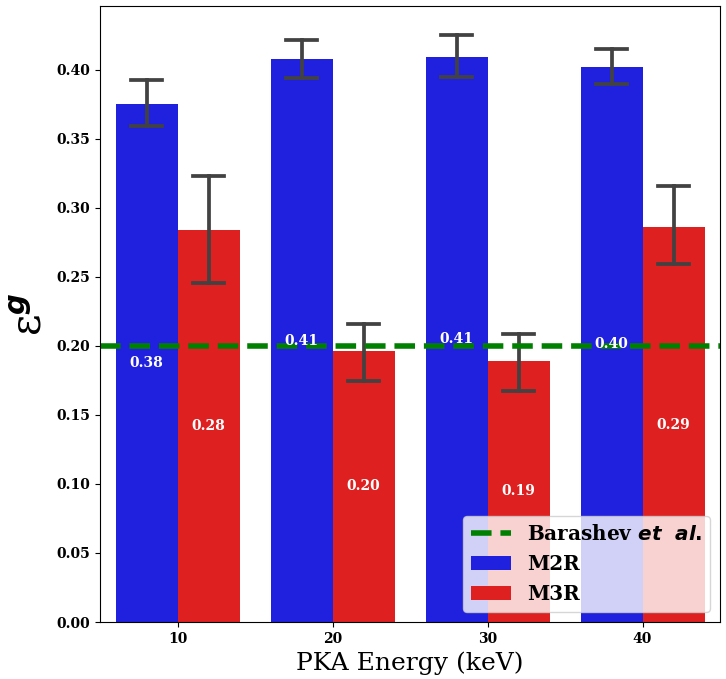}
                \vspace{-2em}
            \end{subfigure}%\\[\baselineskip]
            \begin{subfigure}{\textwidth}
                \centering
                \caption*{(d)}
                \includegraphics[width=\textwidth]{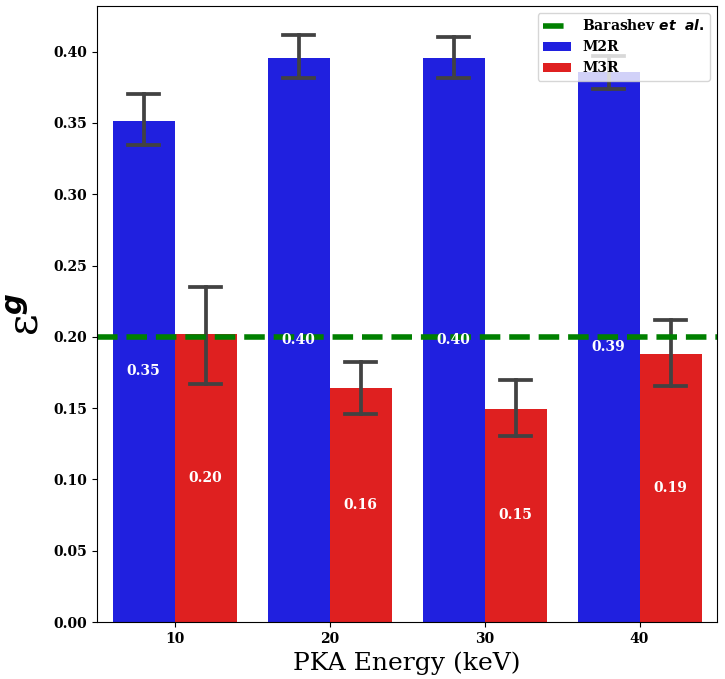}
                \vspace{-2em}
            \end{subfigure}%\\[\baselineskip]
            % \caption*{Column 2}
            \label{fig:col2}
        \end{minipage}
    \caption{The effect of minimum cluster size and cluster type on $\varepsilon_{i}^{g}$ based on automatic analysis}
    \label{fig:Epsilon}
\end{figure*}

\subsection{Automated analysis }
Our automated analysis involved a more permissive definition of ADC than visual analysis. Indeed, automated analysis allows for a cluster to move in all three dimensions and be classified as an ADC, if $D_{a}$ is considerably larger than $D_{c}$. We verified the sensitivity of our assessment to this ratio and tested $D_{a}/D_{cl}$ cutoffs ranging between 4 and 10 (see \cref{fig:DcDaAutomated}). This choice has a minor impact on the estimated $\varepsilon_{i}^{g}$.

$\varepsilon_{i}^{g}$ values estimated using different $D_{a}/D_{cl}$ ratios and minimum accepted cluster sizes are reported in \cref{fig:Epsilon}. This study is concluded on the minimum cluster size of 3 and $D_{a}/D_{cl}=4$. The automated analysis and visual analysis lead to a similar estimate of $\varepsilon_{i}^{g}$ ($\sim$24\%) when considering M3R. Visual inspection and the automated analysis led to different estimates of $\varepsilon_{i}^{g}$ when M2R was used ($\sim$19\% vs $\sim$39\%). The automated analysis slightly tends to give larger estimates of $\varepsilon_{i}^{g}$ compared to visual analysis. As explained above, our automated analysis involved a more permissive definition of ADC than visual analysis. In fact, automated analysis allows for a cluster to move in all three dimensions and still be classified as an ADC, if D$_a$ is considerably larger than D$_c$.

% \vspace{-1em} 

\section{Limited annealing}
\begin{figure}[H]
\centering
\includegraphics[width=0.9\linewidth]{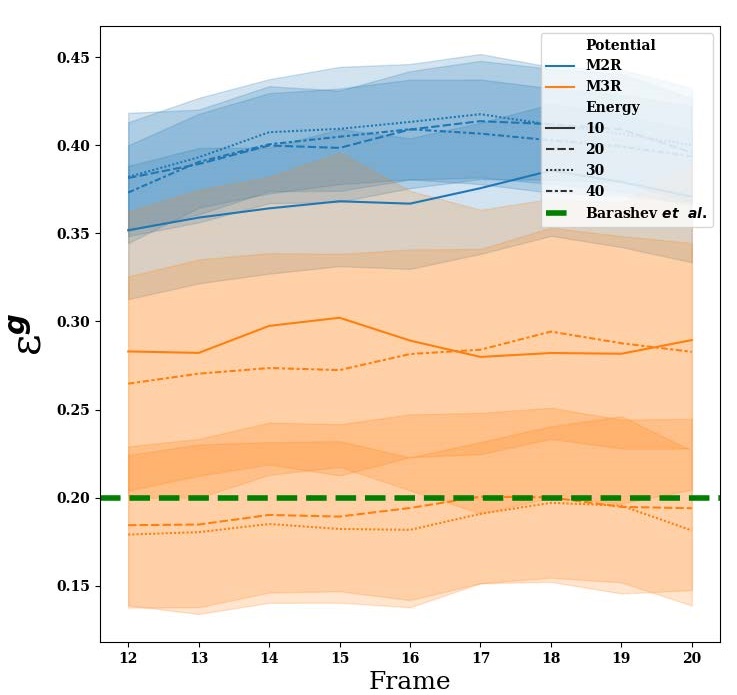}
% \vspace{-1em} 
\caption{Effect of annealing time on $\varepsilon_{i}^{g}$ of cluster containing 3 or more SIAs.}
\label{fig:Annealing}
\end{figure}

As the cascade debris ages, defects can either merge or annihilate (in a real system, they can also interact with other sinks, which are not captured by our MD simulation). The effect of this limited annealing on $\varepsilon_{i}^{g}$ is illustrated in \cref{fig:Annealing}. We opted to calculate $\varepsilon_{i}^{g}$ based on the remaining debris at the final timestep.

\section{Equivalent sphere analysis}
\begin{figure}[H]
\centering
\includegraphics[width=0.65\linewidth]{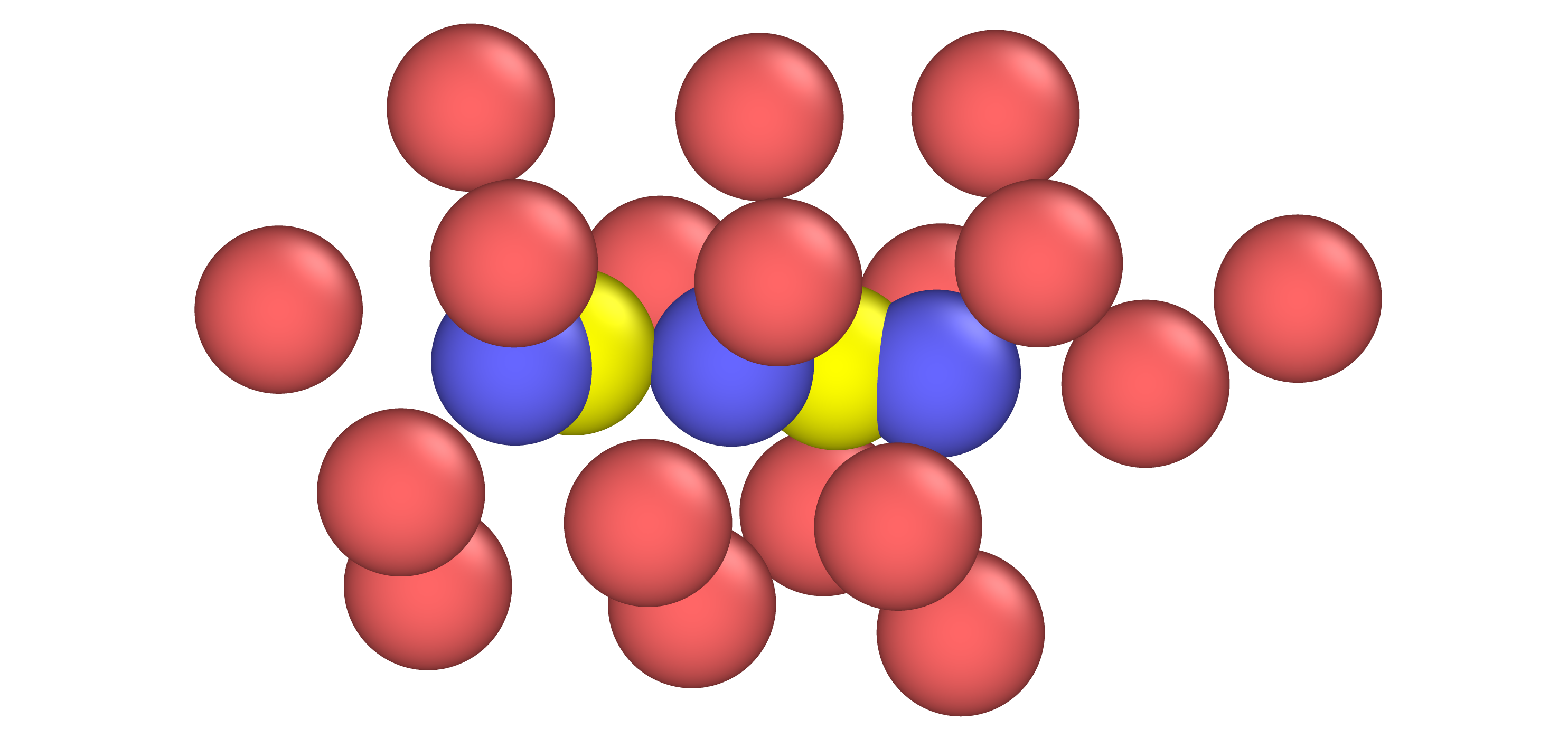}
% \vspace{-1em} 
\caption{A single SIA visualized using equivalent sphere analysis in perspective view.}
\label{fig:Esa}
\end{figure}

The structure of 1-D gliding clusters was visualized using equivalent sphere analysis. By comparing each configuration with the reference perfect structure, the missing positions are assigned as vacancies in this method. Similarly, surplus-filled positions where atomic distance is more than vibration allowance (D$_V$ = 0.2 D$_{Eq}$) and less than neighbouring allowance (D$_N$ = D$_{Eq}$ - 2 D$_V$) are considered as SIAs. The contradistinction of SIAs and vacancies determines the type of defect cluster. For example, \cref{fig:Esa} depicts a single SIA where blue, yellow, and red balls represent 3 SIAs, 2 vacancies and some distorted atoms, respectively.

% \vspace{-2em} 

\section{Self-interstitial atom cluster bias}
\begin{figure*}
\centering
\includegraphics[width=0.5\linewidth]{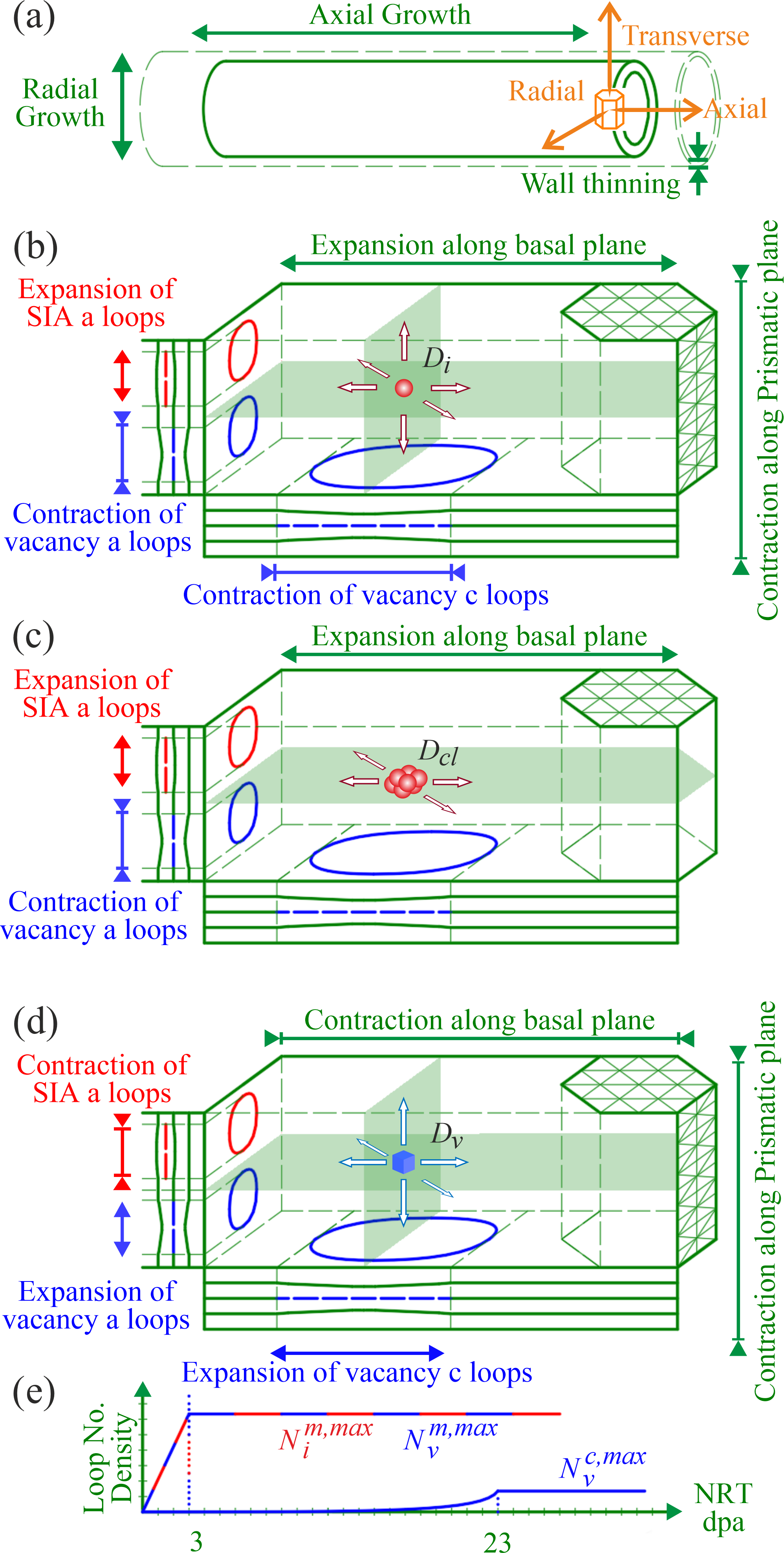}
\caption{Schematic representation of SIACB model. In this model, absorption of defects by sinks, i.e. loops, results in RIG \cite{barashev2015theoretical}. (a) Under high-energy neutron fluence, the pressure tube undergoes dimensional changes such as axial elongation, diametral expansion and wall thinning \cite{causey1988effect}. The pressure tube is highly textured and the majority of basal plane normals are in the transverse direction \cite{holt1979factors}. (b) isotropic diffusion of single SIAs ends in the expansion of SIA-type a-loops and the contraction of vacancy-type a and c-loops. These in return contribute to the expansion along the basal plane and contraction along the prismatic plane. (c) The basal glide of SIA clusters has a similar effect on the a-loops while the c-loops are intact. Hence, their preferential glide contributes to the expansion along the basal plane. (d) single vacancies diffuse isotropically and have an opposite effect on loops compared to single SIAs, i.e. the expansion of a and c-type vacancy loops and contraction of SIA-type the a-loops. This in return causes contraction along prismatic and basal directions. (e) In this model, the a-loops are saturated at early irradiation doses ($\sim$ 3 NRT dpa) while c-loops are saturated later ($\sim$ 23 NRT dpa) at one order of magnitude lower loop number density.}
\label{fig:SchematicSiacb}
\end{figure*}

Barashev, Golubov and Stoller \cite{barashev2015theoretical} formulated the self-interstitial atom cluster bias (SIACB) model by making two main assumptions. First, three diffusing agents contribute to the progress of the radiation induced growth (RIG) strain (\cref{fig:SchematicSiacb} a): single SIAs, 1-D diffusing SIA clusters, and single vacancies. Diffusion of these three agents adds prismatic planes and removes basal planes which contribute to expansion along the basal plane and contraction along the prismatic plane, respectively, as shown schematically in \cref{fig:SchematicSiacb} (b to d).    \Cref{GolubovVacancy,GolubovSia,GolubovCluster} describe the mass balance of these three agents by considering the generation rate, and removal rate at the sinks. Here, $D$ is the diffusion coefficient, $\rho$ is the density of dislocations, $k^2$ is the sink rate, and $G$ is the point defect production after the recombination in the cooling down stage. In these equations, single vacancies, single SIAs and SIA clusters are represented by $v$, $i$, and $cl$ subscripts, respectively. In this model, $\varepsilon_i^g$ depicts the ratio of SIAs inside irradiation-induced 1D gliding clusters to total produced SIAs and $n_i^g$ shows the mean number SIAs in these clusters. The second assumption is the early saturation of the a-loops and the sudden appearance of vacancy c-loops at an arbitrary dpa which marks the beginning of stage III of RIG, also known as break-away growth (depicted in \cref{fig:SchematicSiacb} e).

Based on these two assumptions, the strain rate is measured by accounting for the interaction of dislocation loops and the flux (net number) of SIA defects, i.e. SIA clusters along basal directions ($\frac{1}{3}\chi G^{\text{NRT}}$), and single SIAs along basal and prismatic directions ($-\rho_{c,a_i}(D_{v}C_{v}-D_{i}C_{i})$). \Cref{StrainRateBasal,StrainRatePrismatic} show this dependency of strain rate on the effective dislocation densities, where $\phi$ is $\phi = G^{\text{NRT}}t$ is the irradiation dose in dpa, $G^{\text{NRT}}$ is the Norgett-Robinson-Torrens (NRT) standard dose rate (dpa per second), and $\chi=(1-\varepsilon_r)\varepsilon_i^g$ represents the fraction of the total displaced atoms formed in 1-D gliders. Integration over the irradiation dose will yield the overall strain along basal and prismatic directions.

\begin{figure}
\centering
\includegraphics[width=0.99\linewidth]{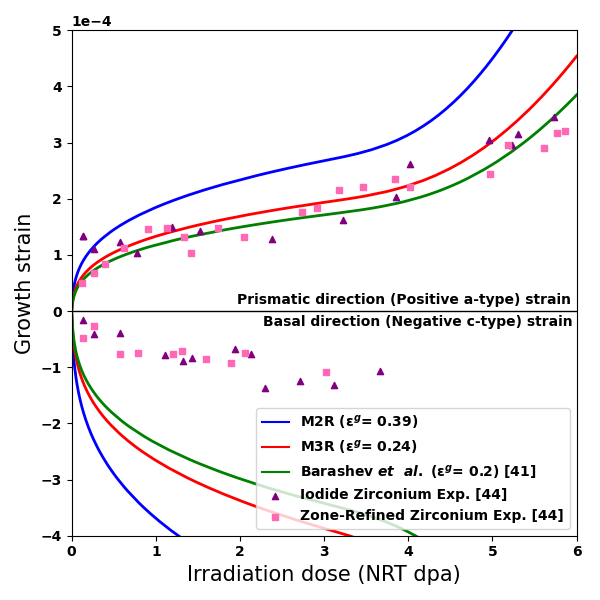}
\caption{SIACB model predicting RIG for $\varepsilon_{i}^{g}$ values measured from MD calculations using M2R and M3R potentials. For comparison, experimental results for RIG in iodized and zone-refined zirconium are also provided \cite{rogerson1988irradiation}.}
\label{fig:Growth}
\end{figure}

We implemented this model (details found here \cite{barashev2015theoretical,golubov2014breakthrough,barashev2012theoretical}) using the measured $\varepsilon_{i}^{g}$ values associated with M2R and M3R. The results are presented in \cref{fig:Growth}. Also, the model predictions were compared to experimental measurements at 553K from \cite{rogerson1988irradiation,carpenter1981irradiation}. \Cref{table:Siacb} describes the parameters used here. The \href{run:/SM/Models/20230525-SIACB.m}{original code} written by the authors \cite{barashev2015theoretical} is also provided, generated to be used in MATLAB \cite{MATLAB}. This model assumes an isotropic distribution of prismatic dislocations ($\rho_{d}^{a_1}$ = $\rho_{d}^{a_2}$ = $\rho_{d}^{a_3}$ = $10^{12}$ $m^{-2}$ ) which is not true in the case of experiments. Therefore, the predicted c-strain (negative strain along prismatic direction) is double that of experimental values \cite{barashev2012theoretical}.

\begin{equation}
\dot{C_{v}}=G - {\rho}D_{v}C_{v}=0,
\label{GolubovVacancy}
\end{equation}

% \vspace{-1em} 

\begin{equation}
\dot{C_{i}}=G(1-\varepsilon_i^g) - {\rho}D_{i}C_{i}=0,
\label{GolubovSia}
\end{equation}

% \vspace{-1em} 

\begin{equation}
\dot{C}_{cl}^{m}=G\frac{\varepsilon_{i}^{g}}{3n_{i}^{g}}-k_{m}^{2}D_{cl}C_{cl}^{m}=0,
\label{GolubovCluster}
\end{equation}

% \vspace{-1em} 

\begin{equation}
\frac{d \varepsilon_{x,y}}{d\phi}=\chi(\frac{1}{2}-\frac{\rho_{x,y}}{\rho}),
\label{StrainRateBasal}
\end{equation}

% \vspace{-1em} 

\begin{equation}
\frac{d \varepsilon_{z}}{d\phi}=-\chi\frac{\rho_{z}}{\rho},
\label{StrainRatePrismatic}
\end{equation}

% \vspace{-1em} 

\begin{table*}
\caption{Details of SIACB model}
\begin{tabular}{lllll}
\multicolumn{1}{c}{Parameter} & \multicolumn{1}{c}{Name} & \multicolumn{1}{c}{Dimensions} & \multicolumn{1}{c}{Value} & \multicolumn{1}{c}{Description}                                                                                                                                   \\ \hline
\multicolumn{5}{l}{\textit{\textbf{Initial conditions}}}                                                                                                                                                                                                                                 \\
$\rho_{d}^{a_1}$                             & rhoda1                   &       $m^{-2}$                         & $10^{12}$                         & \multirow{4}{*}{\begin{tabular}[c]{@{}l@{}}- Densities of edge dislocations with Burgers vectors\\ along a1, a2, a3 and c directions, respectively\end{tabular}} \\
\rho_{d}^{a_2}                             & rhoda2                   &         $m^{-2}$                       & $10^{12}$                         &                                                                                                                                                                  \\
$\rho_{d}^{a_3}$                             & rhoda3                   &       $m^{-2}$                         & $10^{12}$                         &                                                                                                                                                                  \\
$\rho_{d}^{c}$                             & rhodc                    &         $m^{-2}$                       & $0.6\times10^{12}$                         &                                                                                                                                                                  \\
$b_{j}^{a}$                             & bura                     & m                              & $3.0\times10^{-10}$                          & \multirow{2}{*}{- Length of Burgers vectors of a and c dislocations}                                                                                             \\
$b_{j}^{c}$                             & burc                     & m                              & $5.0\times10^{-10}$                         &                                                                                                                 \vspace{2em}                                                  \\ \hline
\multicolumn{5}{l}{\textit{\textbf{Irradiation conditions}}}                                                                                                                                                                                                                             \\
$G^{NRT}$                             & GNRT                     & NRT dpa/s                      & $10^{-7}$                         & - NRT standard dose rate                                                                                                                                         \\
1-\varepsilon_{r}                             & esurv                    & ---                            &   0.1                       & - Defect survival fraction in displacement cascades                                                                                                              \\
\varepsilon_{i}^{g}                             & eica                     & ---                            &                          & \multirow{2}{*}{\begin{tabular}[c]{@{}l@{}}- Fraction of SIAs produced in cascades in the
 \\ form of interstitial glissile clusters\end{tabular}}                        \vspace{2em}                                                                                           \\ \hline
\multicolumn{5}{l}{\textit{\textbf{Reactions and loop nucleation}}}                                                                                                                                                                                                                      \\
$r_{cd}$                            & rdcap                    & m                              & $6.0\times10^{-10}$                        & \multirow{3}{*}{\begin{tabular}[c]{@{}l@{}}- Capture radii of dislocations and interstitial and\\ vacancy loops for glissile SIA clusters\end{tabular}}          \\
$r_{cvl}$                             & ricap                    & m                              & $6.0\times10^{-10}$                        &                                                                                                                                                                  \\
$r_{cil}$                             & rvcap                    & m                              & $6.0\times10^{-10}$                        &                                                                                                                                                                  \\
$\phi_{v}^{max}$                            & phiAI                    & NRT dpa                        & 3.84                        & \multirow{2}{*}{\begin{tabular}[c]{@{}l@{}}- Doses at the end of the nucleation stages of\\ prismatic vacancy and interstitial loops, respectively\end{tabular}} \\
$\phi_{i}^{max}$                            & phiAV                    & NRT dpa                        & 3.84                        &                                                                                                                                                                  \\
$N_{v}^{m,max}$                            & dvlam                    &         $m^{-3}$                       & $3.3\times10^{21}$                        & \multirow{2}{*}{\begin{tabular}[c]{@{}l@{}}- Max. values of the number density of vacancy and \\ interstitial prismatic sessile loops\end{tabular}}              \\
$N_{i}^{m,max}$                            & dilam                    &         $m^{-3}$                      & $3.3\times10^{21}$                        &                                                                                                                                                                  \\
$\phi_{v_0}^{c}$                             & phiC                     & NRT dpa                        & 3                        & - Starting dose of vacancy c–type loops formation                                                                                                                \\
$\Delta$                            & delta                    & NRT dpa                        & 20                        & - Ending dose of the nucleation of vacancy basal sessile loops                                                                                                   \\
$A$                            & AA                       & ---                            & 5                        & - Parameter of nucleation of vacancy basal sessile loops                                                                                             \\
$N_{V}^{c,max}$                            & dvlcm                    &         $m^{-3}$                       & $1.0\times10^{21}$                        & - Max. number of density of vacancy basal sessile loops                                                     \vspace{2em}                                           \\ \hline
\multicolumn{5}{l}{\textit{\textbf{Integration and output}}}                                                                                                                                                                                                                             \\
$\phi^{max}$                             & phimax                   & NRT dpa                        & 40                        & - Max. dose                                                                                                                                                      \\
$\delta\phi$                            & dphi                     & NRT dpa                        & $10^{-4}$                        & - Dose integration step                                                                                                                                                                                 
\end{tabular}
\label{table:Siacb}
\end{table*}

\clearpage
\bibliographystyle{Setup/elsarticle-num.bst} 
\bibliography{Manuscript.bib}
% \bibliographystyle{unsrt}
% \bibliography{xampl}
% \end{multicols}